\documentclass[fleqn,usenatbib]{mnras}

\usepackage{multirow}
\usepackage{caption}
\usepackage{soul} 
\usepackage{color, xcolor} 
\usepackage{graphicx}  
\usepackage{subcaption} 
\usepackage{booktabs}  
\usepackage{multirow}  
\usepackage{tabularx}

\usepackage{ulem} 

\usepackage{newtxtext,newtxmath}

\usepackage[T1]{fontenc}

\DeclareRobustCommand{\VAN}[3]{#2}
\let\VANthebibliography\thebibliography
\def\thebibliography{\DeclareRobustCommand{\VAN}[3]{##3}\VANthebibliography}

\usepackage{graphicx}	
\usepackage{amsmath}	

\title[Rapid automatic detection method of moving objects]{Rapid Automatic Multiple Moving Objects Detection Method Based on Feature Extraction from Images with Non-sidereal Tracking}

\author[Lei Wang et al.]{
Lei Wang,$^{1,2}$\thanks{E-mail: wanglei@bao.ac.cn}
Xiaoming Zhang,$^{1,2}$\thanks{E-mail: xiaomingzhang@bao.ac.cn}
Chunhai Bai,$^{3}$
Haiwen Xie,$^{2,4}$
Juan Li,$^{1,2}$\newauthor
Jiayi Ge,$^{1,2}$
Jianfeng Wang,$^{1,2}$
Xianqun Zeng,$^{1}$
Jiantao Sun,$^{3}$
and Xiaojun Jiang$^{1,2}$\thanks{E-mail: xjjiang@bao.ac.cn}
\\
$^{1}$CAS Key Laboratory of Optical Astronomy, National Astronomical Observatories, Chinese Academy of Sciences, Beijing 100101, China\\
$^{2}$University of Chinese Academy of Sciences, Beijing 100049, China\\
$^{3}$Xinjiang Astronomical Observatory,Chinese Academy of Sciences, Urumqi 830011, China \\
$^{4}$Changchun Observatory, National Astronomical Observatories, Chinese Academy of Sciences, Changchun 130117, China 
}

\date{Accepted 2024 September 02. Received 2024 August 28; in original form 2024 June 26}

\pubyear{2024}

\begin{document}
\label{firstpage}
\pagerange{\pageref{firstpage}--\pageref{lastpage}}
\maketitle

\begin{abstract}

Optically observing and monitoring moving objects, both natural and artificial, is important to human space security. Non-sidereal tracking can improve the system’s limiting magnitude for moving objects, which benefits the surveillance.
However, images with non-sidereal tracking include complex background, as well as objects with different brightness and moving mode, posing a significant challenge for accurate multi-object detection in such images, especially in wide field of view (WFOV) telescope images.
To achieve a higher detection precision in a higher speed, we proposed a novel object detection method, which combines the source feature extraction and the neural network. First, our method extracts object features from optical images such as centroid, shape, and flux. Then it conducts a naive labeling based on those features to distinguish moving objects from stars. After balancing the labeled data, we employ it to train a neural network aimed at creating a classification model for point-like and streak-like objects. Ultimately, based on the neural network model's classification outcomes, moving objects whose motion modes consistent with the tracked objects are detected via track association, while objects with different motion modes are detected using morphological statistics. The validation, based on the space objects images captured in target tracking mode with the 1-meter telescope at Nanshan, Xinjiang Astronomical Observatory, demonstrates that our method achieves 94.72\% detection accuracy with merely 5.02\% false alarm rate, and a processing time of 0.66s per frame. Consequently, our method can rapidly and accurately detect objects with different motion modes from wide-field images with non-sidereal tracking.

\end{abstract}




\section{Introduction}

With the increasing emphasis on planetary defense missions and human space activities, the monitoring of natural celestial bodies (asteroids, meteors, etc.) or space objects (satellites, space debris, etc.) in near-earth space are of great significance to the safety of human space activities \citep{2017spde.confE.220V}. Non-sidereal tracking is a common observation method, which can improve the observation accuracy and signal-to-noise ratio (SNR) of moving targets.
However, under the observation condition of non-sidereal tracking, especially in WFOV telescope images, there may be multiple moving objects in a single field of view in the observed image, and objects move at a different speed and direction \citep{2016AdSpR..57.1607V}. 
How to quickly detect multiple moving objects in wide-field images with non-sidereal tracking is an urgent technical challenge to be solved.

Traditional image processing methods for detecting moving objects include image stacking \citep{2012amos.confE..13Y}, line identification method \citep{zhang2024detecting}, multi-stage hypothesis testing (MHT) method \citep{li2019space,liu2020space,du2022dp}, and methods based on 3D matched filtering \citep{reed1983application}. The image stacking method has significant advantages in detecting faint and weak targets in small fields of view. However, under a dense star background in a large field of view, stars can stack into long streaks, thereby affecting the accuracy of target detection. Line identification methods can extract the track of the point target moving in a straight line, but under the condition of low SNR, it is easy to be drowned by noise and has poor anti-interference and real-time performance. The MHT method needs to search the prior information of all linear tracks or moving objects in the input image, which is high in computational cost and long in analysis time for multi-target detection and is not suitable for real-time detection. 3D matched filters can detect slow-moving point targets, but the parameters of the filter need to be set in advance.

In recent years, many deep learning methods have been used for the detection of moving objects, such as Xi et al. \citep{xi2020space} applied the convolutional neural network model (CNN) of LeNet5 architecture, which was trained using a large number of preprocessed simulated streaks with different SNR and motion parameters, effectively detecting low SNR space debris streaks. Wang et al. \citep{wang2022discovering}  trained a convolutional neural network using artificial streaks generated by simulation, and then detected near-Earth celestial objects. Tao et al. \citep{tao2023sdebrisnet} used convolutional neural networks to add an attention mechanism to calculate the centroid coordinates of space debris from satellite video sequences. Although deep learning methods can accurately detect moving objects, they usually require a large number of optical images as training data, and using only simulated data often yields less effective results, especially for space objects against complex backgrounds. Deep network models typically struggle to accurately construct appearance models, particularly for complex background environments and moving objects with different parameters that have not appeared in training data, leading to the overall robustness and generality of current deep learning methods not being high \citep{jia2020detection}. However, with the advancement of deep learning methods, there is significant potential and prospects for the application of deep learning methods in the detection of moving objects.

Many researchers have been committed to the study of moving target detection issues, but in actual observations, factors such as complex sky backgrounds, dense stars, uncertain telescope pointing, non-uniform imaging systems, and optical system distortions affect the precision and accuracy of moving objects detection \citep{sun2019algorithms}. Especially for multiple moving objects, each possessing distinct motion modes such as speed and direction, it becomes challenging to simultaneously optimize the accuracy rate and minimize the false alarm rate of the target detection method. Additionally, the timeliness and robustness of the method must be taken into account. At present, it is still an urgent challenge to detect multiple moving objects quickly and accurately from wide-field images with non-sidereal tracking.

In non-sidereal tracking mode, the sensor remains pointed at the moving target. Thus, the detected moving target appears as point and stars appear as streaks. The non-sidereal tracking mode maintains the stability of a moving target within the field of view, thereby reducing image blur caused by the target's movement. Compared to the sidereal tracking mode, this improves the SNR of the target in the image \citep{sharma2023astreaks}. In non-sidereal tracking images, the dense stellar streaks can affect the detection of targets. Over the years, various target detection methods for non-sidereal tracking images have been proposed. Pan et al. enhanced the SNR and implemented adaptive background noise suppression to achieve target detection \citep{pan2014detection}. Sun et al. proposed a point target extraction algorithm based on mathematical morphological processing and multi-frame median filtering \citep{sun2015algorithms}. Jiang et al. removed stars by applying wavelet transform and guided filtering to three consecutive frames, and then used robust principal component analysis to separate the targets from the background in single-frame images \citep{jiang2022faint}. Lin et al. proposed a space target detection algorithm that utilizes the distribution characteristics of the targets \citep{lin2021robust}. Although the above methods can achieve the detection of targets, the accuracy of the algorithms is sensitive to the set threshold parameters, and they do not take into account other moving objects with different motion patterns in the image. Currently, there are also many target detection methods based on deep learning. For example, Dai et al. proposed an target detection method based on PP-YOLOv2 \citep{dai2022effective}, Abay et al. put forward the GEO-FPN method based on segmentation ideas \citep{abay2021geo},  and Liu et al. introduced a multi-frame temporal dense nested attention (MT-DNA) method \citep{liu2024multi}. Although these neural network methods have achieved good results, they require a large amount of training data and have high computational costs.

Based on the above challenges, this paper proposes a multiple moving objects detection method based on feature extraction from optical images. The structure of this paper is as follows: Section 2 gives the details of our algorithm; Section 3 provides the experimental results and analysis; Section 4 draws conclusions and prospects.



\section{Methods}

When detecting moving objects with optical telescopes, the images can be regarded as the superimposed result of moving objects, star, background, and noise \citep{2022PASP..134b4503J}. The modeling formula is: 
\begin{equation}
f(x,y,k)=O(x,y,k)+S(x,y,k)+B(x,y,k)+N(x,y,k) 
\textbf{\label{eq:1}}
\end{equation}

Where (x,y) represents the pixel position in the image. For an image with a size of m×n, $x$=1,2…m, $y$=1,2…n, $k$ represents the frame number in the image sequence, $f(x,y,k)$ represents the brightness value of the image pixel, $O(x,y,k)$ represents the target brightness value, $S(x,y,k)$ represents the brightness value of stars, $B(x,y,k)$ represents the brightness value of the background, and $N(x,y,k)$ represents the sum of the brightness values of various noise.

\begin{figure*}
	\includegraphics[width=2\columnwidth]{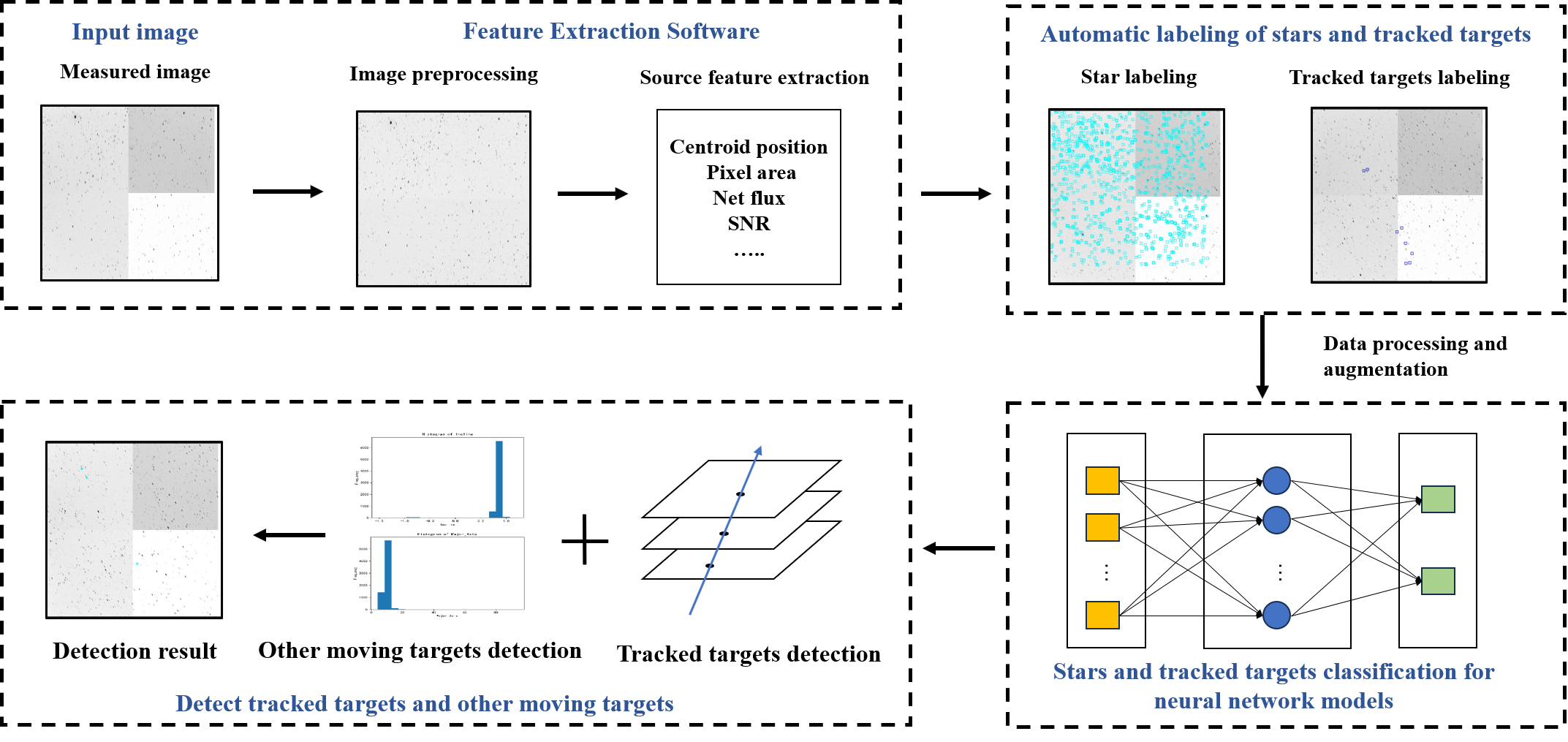}
    \caption{Algorithm flow of this article. To clearly display the sources in the image, we performed a black-and-white color inversion on the image. }
    \label{fig:fig1}
\end{figure*}

As mentioned above, the photon count value of each pixel in the image is a combination of background signal, source signal (stars and moving objects), and noise signal. The experimental images used in this paper are from space object images observed by Nanshan One-meter Wide-field Telescope \citep{bai2020wide}, configured in target tracking mode where stars appear as streaks and space object as points. The structure of the moving objects detection algorithm proposed in this paper (as shown in Fig.~\ref{fig:fig1}). 
Firstly, image preprocessing and feature extraction are carried out to accurately extract the source and its feature information whose brightness is greater than the threshold value. Then, we use image registration methods, clustering statistics, and track association methods to achieve classification of stars and tracked targets, significantly reducing the time cost of manual annotation. After obtaining the classification results, we optimize the annotated data through data balancing and input them into a well-designed neural network model for training, aiming to achieve accurate classification of stellar and tracked target features. Finally, by combining the classification results of the neural network model and the motion track association method, we effectively eliminate false alarms and accurately detect the tracked targets. Furthermore, by integrating the classification results of the neural network model and the morphological statistical information in each frame, we are able to detect moving objects with different speeds and directions.


\subsection{Source detection and feature extraction}
\label{sec:Image feature extraction pipeline} 

\begin{figure}
\centering
	\includegraphics[width=.6\columnwidth]{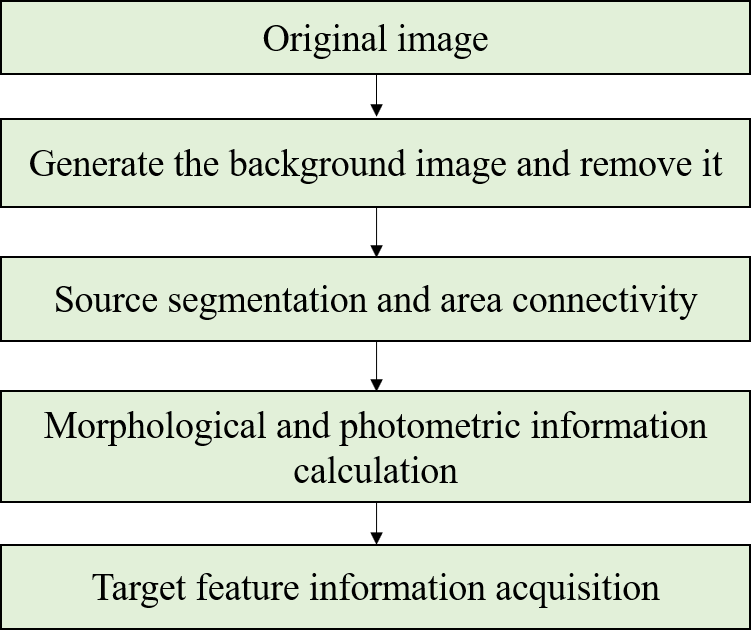}
    \caption{Feature extraction from images process in Feature Extraction Software. }
    \label{fig:fig2}
\end{figure}

Feature Extraction Software is designed to realize source detection and feature extraction. Source detection and feature extraction process mainly includes the image's background suppression, source segmentation and extraction, calculation of the centroid positions of sources, and flux measurement. The process of source extraction in this paper is shown in Fig.~\ref{fig:fig2}.

\subsubsection{Background suppression}

The non-uniformity in image background is primarily caused by the inconsistency in detector pixel responses and stray light. This inconsistency interferes with the effective differentiation between sources and background, leading to an increased false alarm rate. Therefore, it is necessary to effectively suppress the image background before extracting the sources. Due to various factors, such as moonlight, clouds, and other influences, the brightness of the background in an image may exhibit unevenness in different regions. Consequently, segmenting the image into multiple small local regions is crucial for accurately estimating the local background in astronomical images when generating background images.
Within each local region, we implement a process to eliminate pixel values outside the median ± 3$\sigma$ range \citep{stetson1987daophot}, iteratively refining until all values are within this range. Subsequently, a background image is generated through two-dimensional cubic spline interpolation on the filtered areas. Finally, the background image is subtracted from the original image.
To mitigate the adverse effects of noise in images after background subtraction, we adopt median filtering technology to reduce noise. This step will enhance the accuracy of subsequent feature extraction and classification.

\subsubsection{ Source segmentation and extraction}

After reducing the impact of background and noise in the image, source segmentation is executed through an image segmentation technique predicated on gray-scale thresholding. This method describes pixels into identical categories based on their luminance values, by establishing a segmentation threshold relative to the local area's background image. Consequently, entities with distinct brightness are differentiated from the backdrop, and the segmentation of sources follows the following rules:

(1) The extracted pixel value after deducting the background is higher than the detection threshold $T_{\text{seg}}$, and the threshold $T_{\text{seg}}$ is calculated as shown in equation~(\ref{eq:2}).

(2) Perform 8-neighbor connectivity labeling on the detected pixels.

(3) Calculate and extract the area of the connected region of pixels.

(4) Filter out the connected domain whose area is less than the minimum area $Area_{min}$.

(5) Filter out the connected domain whose area is smaller than the connected domain and whose area is larger than the maximum area $Area_{max}$.

In numerous experimental observations, we have discovered that when the area of an image region is less than 3 pixels or exceeds 600 pixels, these regions are likely to be attributed to noise or background interference rather than the stars or targets of our interest. Consequently, we have set the minimum area threshold to 3 pixels ($Area_{\text{min}}$ = 3 pixels) and the maximum area threshold to 600 pixels ($Area_{\text{max}}$ = 600 pixels), excluding those regions that are too small or too large. This approach enhances the accuracy of source extraction.


\begin{equation}
T_{seg}=mean+ k\sigma_b
\textbf{\label{eq:2}}
\end{equation}

Where the $mean$ is the average value of the background image, $\sigma_{b}$ is the standard deviation of the background image, and $k$ is a chosen coefficient. Coefficient $k$ has a great influence on the extraction of the target. If $k$ is too large, it will cause the weak source of the image cannot be extracted, and when k is too small, it may cause more noise to be extracted. After conducting multiple experiments, we found that when the coefficient $k$ equals 2.5, more accurate source information can be obtained, and more weak sources have been extracted. Therefore, in this paper, the coefficient $k$ = 2.5.

\begin{figure}
\centering
	\includegraphics[width=\columnwidth]{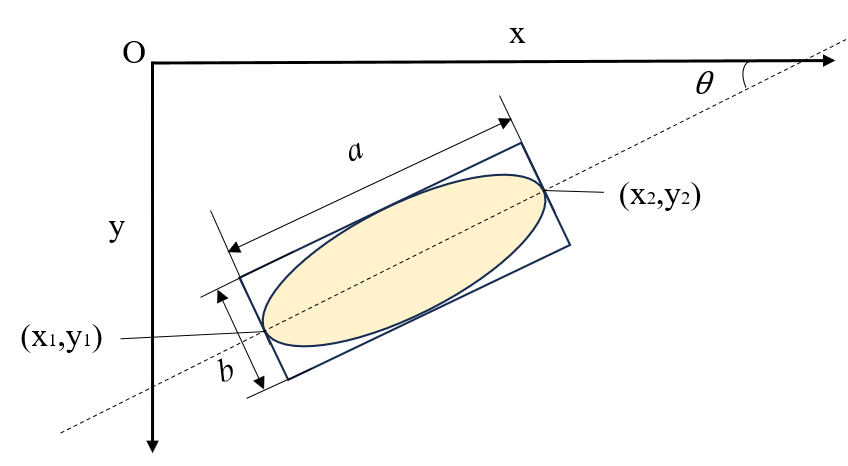}
    \caption{Morphological information of the connected regions. where O is the origin of the pixel coordinate system. $a$ is the major axis length, $b$ is the minor axis length. $\theta$ is the inclination angle, the sign is positive clockwise along the X-axis, negative otherwise. ($x_{1}$, $y_{1}$) and ($x_{2}$, $y_{2}$) are the coordinates of the two endpoints.}
    \label{fig:fig_angle}
\end{figure}

After identifying the connected regions, the maximum span of a connected region is considered as its length, which can be obtained by calculating the farthest distance of the boundary points in a certain direction. This maximum span in that direction is thus taken as the major axis length of the connected region. The maximum span of the connected region in the direction perpendicular to its length is considered as the minor axis length, as shown in the figure~\ref{fig:fig_angle}.

The formula for calculating the circularity C of the contour of the connected region is as equation~(\ref{eq:3}):

\begin{equation}
C = \frac{b}{a}
\textbf{\label{eq:3}}
\end{equation}

Where $a$ is the major axis length, and $b$ is the minor axis length. When the target area is circular, the circularity is the maximum value 1.

The inclination angle $\theta$ of the connected domain is calculated as equation~(\ref{eq:4}):

\begin{equation}
\theta =\pm \arctan\left(\frac{y_{2}-y_{1}}{x_{2}-x_{1}}\right)  
\textbf{\label{eq:4}}
\end{equation}

Where ($x_{1}$, $y_{1}$) and ($x_{2}$, $y_{2}$) are the coordinates of the two endpoints of the connected domain in the direction parallel to the major axis.
And the sign of the inclination angle is defined as positive when it is in the clockwise direction along the X-axis, and negative when in the opposite direction, in units of rad.

\subsubsection{Source centroid position calculation and flux measurement}

The centroid method \citep{Horn} (See equation~(\ref{eq:5})) is used to calculate the center of the source and the aperture photometry method \citep{laher2012aperture} to obtain the flux information of the source in this paper. At the same time, luminance information such as instrument magnitude and background luminance are calculated. We calculate the SNR according to equation~(\ref{eq:6}).

\begin{equation}
(x_0, y_0) = \frac{\left( \sum_{(x,y)\in S} (x,y)I(x,y) \right)}{\left( \sum_{(x,y)\in S} I(x,y) \right)}
\textbf{\label{eq:5}}
\end{equation}

\begin{equation}
SNR= \frac {F \times \sqrt {S}}{\sigma_b }
\textbf{\label{eq:6}}
\end{equation}

Where $(x_0, y_0)$ is the centroid of source, $S$ is source area after segmentation, $I(x, y)$ is the brightness of the original image at $(x,y)$, $F$ is the flux of the source, $\sigma_{b}$ is the standard deviation of the background.

\subsubsection{Feature Extraction Software}

We have integrated the above algorithms into a software implemented in C/C++, named Feature Extraction Software, which is used for source detection and feature extraction in images. It is a multi-threaded application, and the external dependencies of this code include libraries (CFITSIO, SOFA). By extracting features from FITs images, the software can obtain information such as source centroid position, pixel area, flux, instrumental magnitude, SNR, maximum flux value, maximum background flux value, major axis length, minor axis length, circularity, and inclination angle. 
Table~\ref{tab:0} provides a detailed overview of the feature data extracted from the software and their respective definitions.

\begin{figure}
\centering
	\includegraphics[width=\columnwidth]{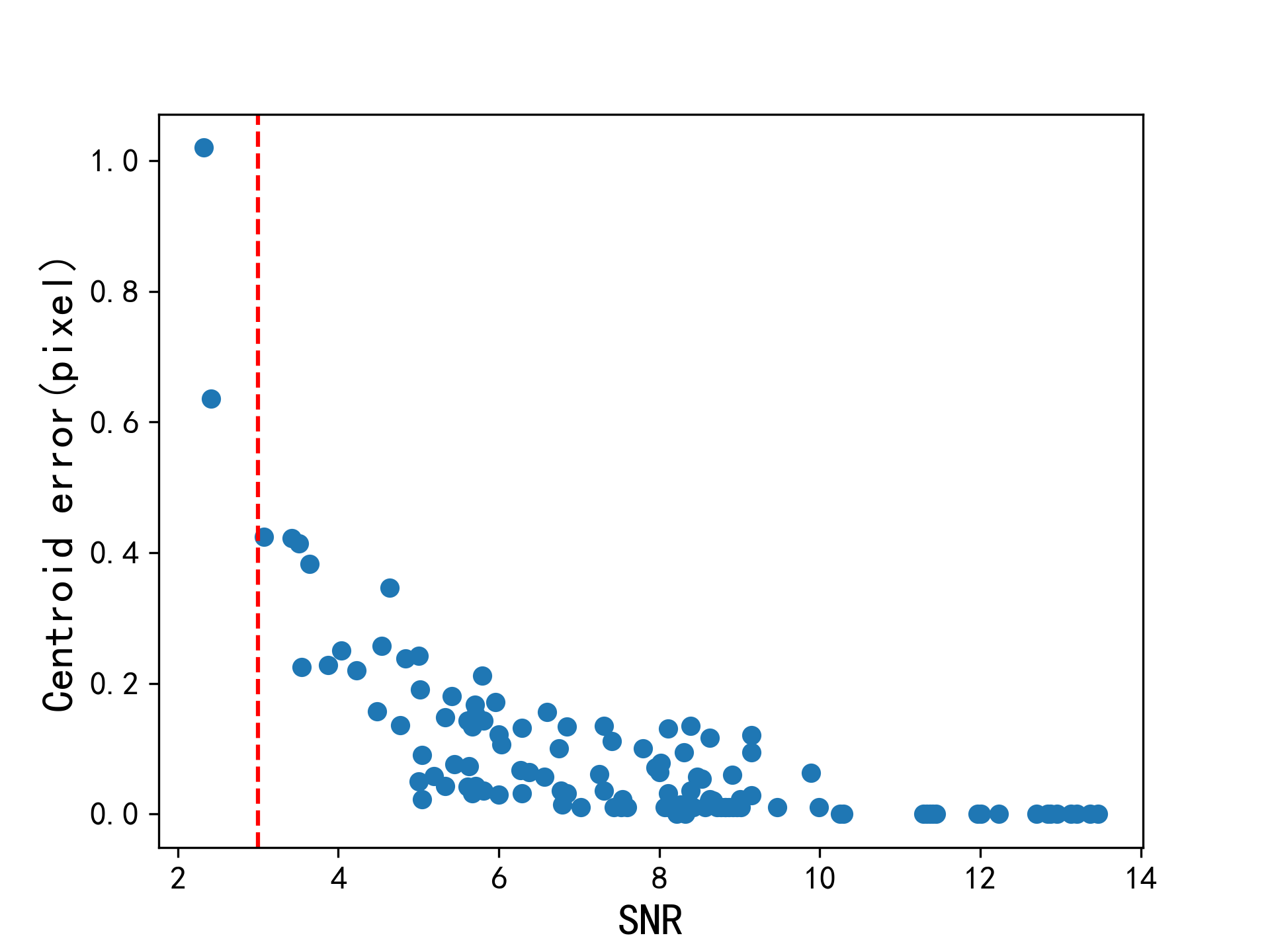}
    \caption{Centroid detection accuracy distribution of simulated targets (FWHM=5 pixels) with different SNR in Feature Extraction Software. The horizontal axis represents the SNR of the source, the vertical axis represents the centroid error extracted by the Feature Extraction Software from the source. The red vertical line represents an SNR of 3.}
    \label{fig:fig3}
\end{figure}

\begin{figure}
\centering
	\includegraphics[width=\columnwidth]{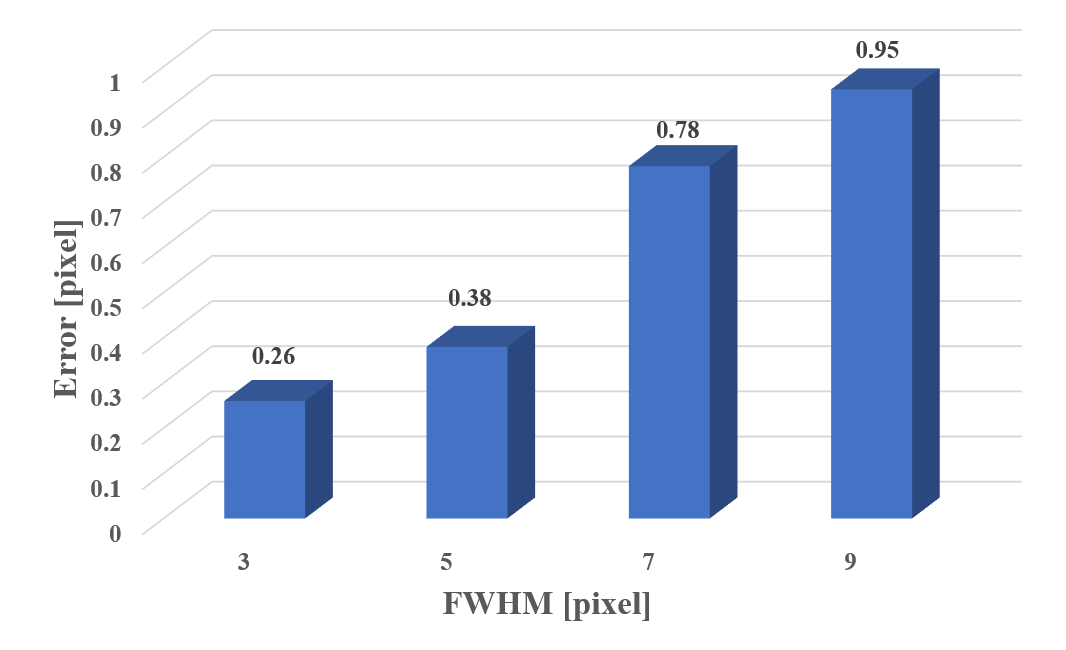}
    \caption{The variation of average centroid error for simulated point targets with different FWHM values, where the SNR of the targets ranges from 3 to 6.}
    \label{fig:fig3_FWHM}
\end{figure}

By simulating point source images with different SNR, we obtained the centroid detection accuracy results of the software for targets with different SNR under the condition that the Full Width at Half Maximum (FWHM) is 5 pixels, as shown in Fig.~\ref{fig:fig3}. To investigate the impact of varying FWHM on centroid localization accuracy, we created a set of simulated images with SNR ranging from 3 to 6 and point targets with FWHM values set to 3, 5, 7, and 9 pixels. For each FWHM setting, 90 frames were generated for testing. After processing these images with software to extract their centroids, we subsequently calculated the average centroid error. Fig.~\ref{fig:fig3_FWHM} displays the variations in average centroid error across different FWHM values. Based on the extraction results, the software is capable of successfully extracting targets with high precision when the target SNR exceeds 3.

\begin{table}
	\centering
	\caption{Features extracted by Feature Extraction Software and their definitions.}
	\label{tab:0}
	\begin{tabular}{cc} 
		\toprule  
    Feature & Definition\\
		\hline

        Frame & The number of frames of the image \\
        Img-x & Center point x coordinates of the source \\
        Img-y & Center point y coordinates of the source \\
        Pixels-N & Pixel area of the source  \\
        Flux & Flux of the source  \\
        INST-mag & Instrument magnitude of the source \\
        SNR & SNR of the source\\
        Pix-flux-peak & Single pixel peak flux within the region of the source\\
        Pix-flux-bk & Source area background single pixel flux\\
        Major  & The major axis length of the source \\
        Minor  & The minor axis length of the source \\
        Circularity  & The circularity of the source \\
        Incline & The inclination angle of the source \\
        
	\bottomrule  
	\end{tabular}
\end{table}

\subsection{Moving object detection}
\label{sec:Image feature extraction pipeline}

\begin{figure}
\centering
	\includegraphics[width=\columnwidth]{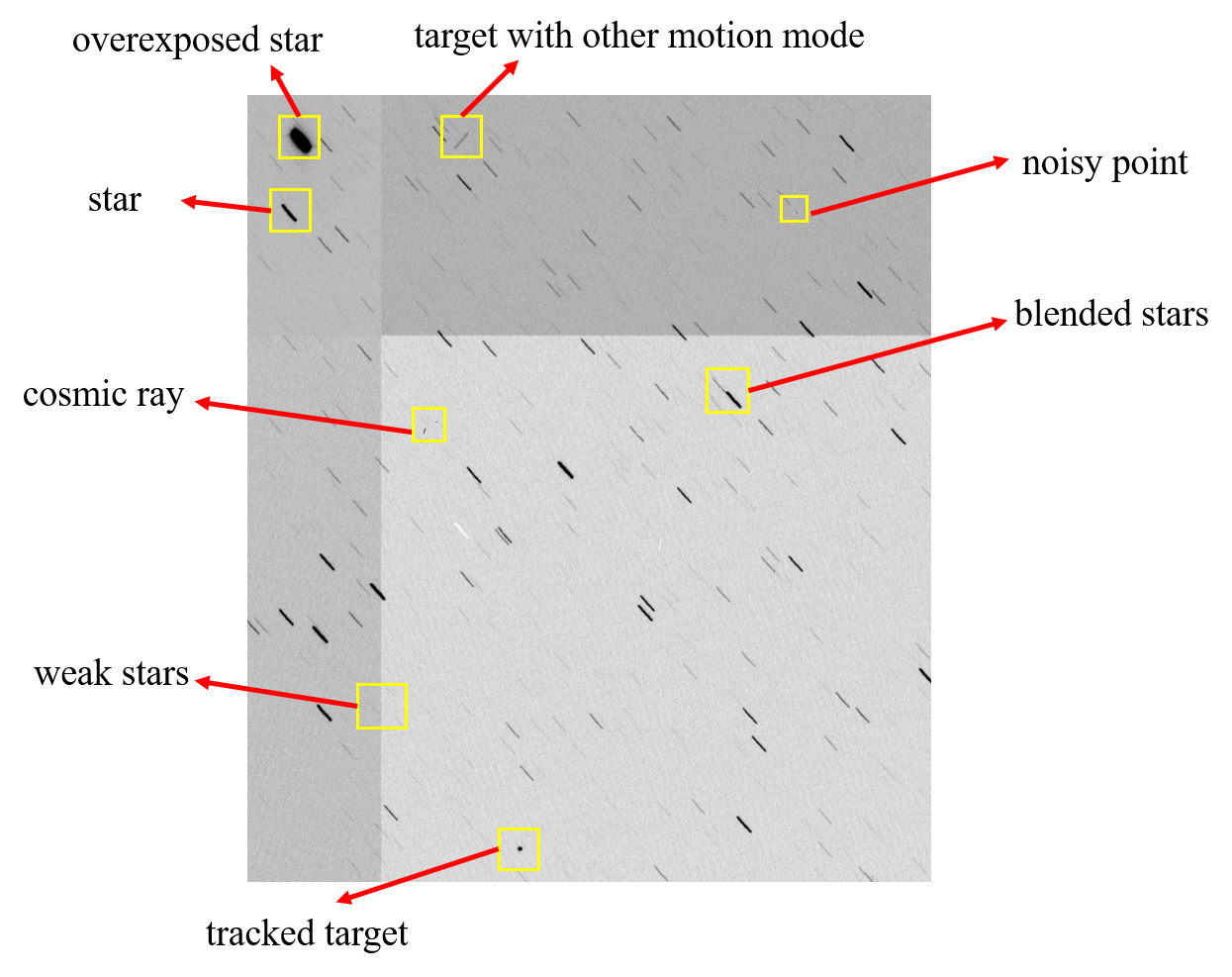}
    \caption{ The types of source extracted from the image include stars (weak stars, blended stars, overexposed stars), moving objects (tracked target, target with other motion mode), noisy points, and cosmic rays, etc. The image is sourced from Nanshan One-meter Wide-field Telescope and has been cropped. We performed a black-and-white color inversion on the image.}
    \label{fig:fig4}
\end{figure}

For images observed by the large field telescope, the source detection using Feature Extraction Software may detect non-stellar and non-target objects. 
Fig.~\ref{fig:fig4} shows all the source types extracted from the images taken by the Nanshan 1-meter Telescope of Xinjiang Astronomical Observatory, including moving objects (tracked targets and targets in other motion modes), stars, cosmic rays, and noise. In the image of target tracking mode, the shape of the source extracted from the image can be divided into two categories, one is striped, such as stars, targets with different motion modes, etc. 
And the other is point-like, such as the tracked target, and the targets consistent with the tracked target motion modes and noise, etc. Cosmic rays in images are usually represented as bright spots, lines or irregular shapes, which are still approximately regarded as point-like or stripe-like here. In order to significantly improve the accuracy and real-time detection of moving objects and reduce the dependence on manually set parameters, this paper adopts a deep learning model to train the multidimensional data of the extraction source, and performs binary classification processing on stars (representing stripy targets) and tracked targets (representing point-like targets). The selection of this classification strategy is based on the fact that these two types of data have high shape representation and obvious features, and the extracted data can be more accurate than random interference factors such as cosmic rays and noise. Classification of these two types of data not only helps us to identify objects with different shapes in the image more accurately, but also provides strong support for the subsequent moving object detection.

Firstly, in order to reduce the time cost of manual labeling, an automatic labeling method is designed to accurately label some stars and tracked targets. Given that the number of stars in the images significantly exceeds the number of tracked targets, we balanced the dataset by generating simulated tracked target images and randomly reducing the number of stars. 
Then, we use the star data to represent the labeled data of the striped object, and the tracked target and the target with the same speed as the labeled data of the point target. By training the deep learning model on these two types of feature data, we get a model that can accurately distinguish the point-like target from the stripy target. In order to further improve the robustness of the system, we use the track association method to remove false alarms such as cosmic rays and noise. By analyzing the motion modes of the tracked target, this method filters out the targets that do not conform to its motion modes, thus reducing the false positive rate. For moving objects whose motion modes are different from tracked targets through the statistical analysis of the morphological modes of the target, we can identify the target that is significantly different from the tracked target and treat it as a potential moving objects for further processing.

\subsubsection{ A naive method for classifying stars and tracked targets}\label{A naive method for classifying stars and tracked targets}

The supervised learning neural network method is adopted to classify tracked targets and stars, which requires input of target label information. To reduce manual operations, this article selects a naive classification method. For the detection of stars, an image registration method is employed, while for tracked targets, a combination of cluster statistics and track association is utilized. In this method, the threshold for associating tracked target tracks is set very low, with the goal of ensuring the accuracy of target detection. 

\paragraph{Star detection based on image registration methods}~{}

The detection of stars can be achieved by utilizing the World Coordinate System (WCS) \citep{thompson2006coordinate} method for astronomical positioning and comparing the data with those in star catalogs. However, when there is no available WCS information, it is necessary to identify stars based on their motion statistical information if one wants to determine the stars in the image. Since all stars move in the same direction and at the same speed in the image, the detection of stars in the image can be effectively achieved through astronomical image registration algorithms \citep{beroiz2020astroalign}. The main steps of star detection using astronomical image registration methods are as follows:

(1)	Search for bright sources within the image, and for these targets, select their four nearest neighbors. Utilize these neighbors to create all possible triangular combinations.

(2)	Establish correspondence between triangles in the images by identifying and matching invariants in both images.

(3)	For each matched pair of triangles, seek the correspondence of vertices and side lengths, thus establishing a point-to-point correlation for the three points.

(4)	Estimate the affine transformation between the two images by finding similar trios of points (triangles) within the images. If this affine transformation aligns with the correspondence of over 80\% of the remaining points (or at least 10 points, whichever is smaller), means the transformation is star motion.

(5)	If the position corresponding to the target in the original image can be found in the transformed image, then it can be determined that the target is a star.

The detection of stars using image registration method can match most of the stars in the image, and the method uses motion statistics, so it is less affected by point spread function and atmospheric conditions.

\paragraph{Object detection based on clustering statistics and Track Association}~{}\label{Object detection based on clustering statistics and Track Association (TA)}

The astronomical image registration algorithm can effectively detect most of the stars in the image. However, due to the challenge of extracting certain faint stars and the fact that stars may exit the field of view as a result of relative motion during target tracking mode, these stars run the risk of being missed during the image registration process. 
We use the clustering method to screen the target candidate for the first time according to the brightness, shape and other information. Then the track association method is used to confirm the second time, and the initial detection of the target is realized. 

K-means algorithm \citep{sinaga2020unsupervised} is a popular unsupervised clustering algorithm, which has the modes of flexibility, high efficiency and easy implementation. The star data detected by the image registration method is removed and the remaining data is classified using K-means. In this paper, K-means classification is carried out by using five characteristic data: pixel area, major axis, minor axis, circularity and inclination angle. For the target tracking images, it can be considered that the category with the longer major axis  length in the morphological features is the category of stars, while the others are considered as target candidates. In order to improve the stability of the algorithm, we use a normalization method to process the data, that is, subtract the minimum value from each data point, and then divide by the difference between the maximum and the minimum value.

\begin{figure}
\centering
	\includegraphics[width=.7\columnwidth]{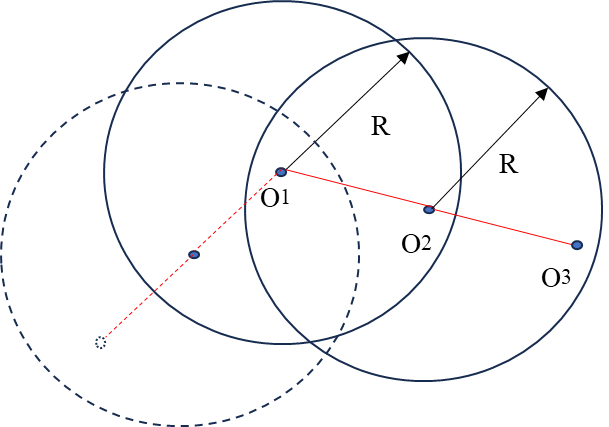}
    \caption{Moving object matching diagram. O1, O2, and O3 are the center positions of the matched target sources. The neighborhood R serves as the search range for target matching. When points in the next frame of the image fall within the R region, they are considered as potential candidate targets. Only when a target is successfully matched in three consecutive frames of images, is it deemed to be the same target across different frames of the image. In target tracking mode, the distance between O1, O2 and O3 is very close.}
    \label{fig:fig5}
\end{figure}

Track association is the process that focuses on the continuity of moving objects' positions across consecutive image frames. Under ideal tracking system conditions, although targets in the physical world may move, due to the high precision of the system, the changes in the pixel positions of these targets between consecutive image frames are usually minimal and can almost be considered static. Therefore, by carefully analyzing these multi-frame position associations, we can significantly differentiate the motion of targets from the background (such as stars).
This study utilizes the KD-Tree method \citep{kubica2007efficient} to associate track of candidate target, and considers the associated data as valid targets. KD-Tree is a data structure widely used in multidimensional search, which can store and quickly retrieve K-dimensional points in the form of a binary tree, and each node can be divided according to different dimensions. This structure can effectively search fast neighborhoods in a massive data environment, and its time complexity is O(nlogn). In this method, we use a neighborhood R, the domain radius is set in the case of the missing object in a few frames, the size of the neighborhood is determined by the modes of the object and the nature of the motion tracking. According to the linear modes of the moving object, only the point that can establish a stable track can be identified as the moving object. That is, if the matching target within the motion range is found in at least three frames of the image, it is considered as a moving object, as shown in Fig.~\ref{fig:fig5}.

The overall process of moving object detection is as follows:

(1) Remove all star data obtained from the image registration method.

(2) Treat the data that do not conform to the star morphology obtained by K-means as target candidates.

(3) The KD-tree method is used to associate track of tracked targets. The neighborhood R, which serves as the search range for target matching, is primarily calculated based on the estimated maximum target motion speed. And 3 consecutive frames of images are identified each time, and the motion track of at least 3 frames of images is associated to determine the moving object.

Based on cluster statistics and track association method, we not only consider the difference of morphology and luminosity between stars and tracked objects, but also associate the motion track of space objects, so we can find most of the point-like moving objects in the image. Based on experimental experience, we set the threshold for associating tracked target tracks to R=10 pixels. Although this setting of R value may cause some tracked targets to be missed, it ensures that there are no false alarms in the labeled tracked targets, thus improving the authenticity of the labeled data.

\subsubsection{Neural network classifies stars and tracked targets}\label{Neural network classifies stars and tracked targets}

Through the method described in Section ~\ref{A naive method for classifying stars and tracked targets}, we successfully utilized a naive approach to generate labeled data for stars and tracked targets. However, this method has the limitation of being cumbersome and time-consuming. In contrast, neural networks, with their end-to-end modes, significantly reduce the need for manual design and intervention, and due to their strong generalization capabilities, they can be applied to observational data from subsequent identical observation modes. Therefore, this paper designs a neural network model and enhances the generalization ability of the network by balancing and augmenting the labeled data obtained through the naive method, thereby achieving intelligent classification of stars and tracked targets.

\paragraph{Neural network design}~{}

\begin{figure*}
\centering
	\includegraphics[width=0.6\textwidth]{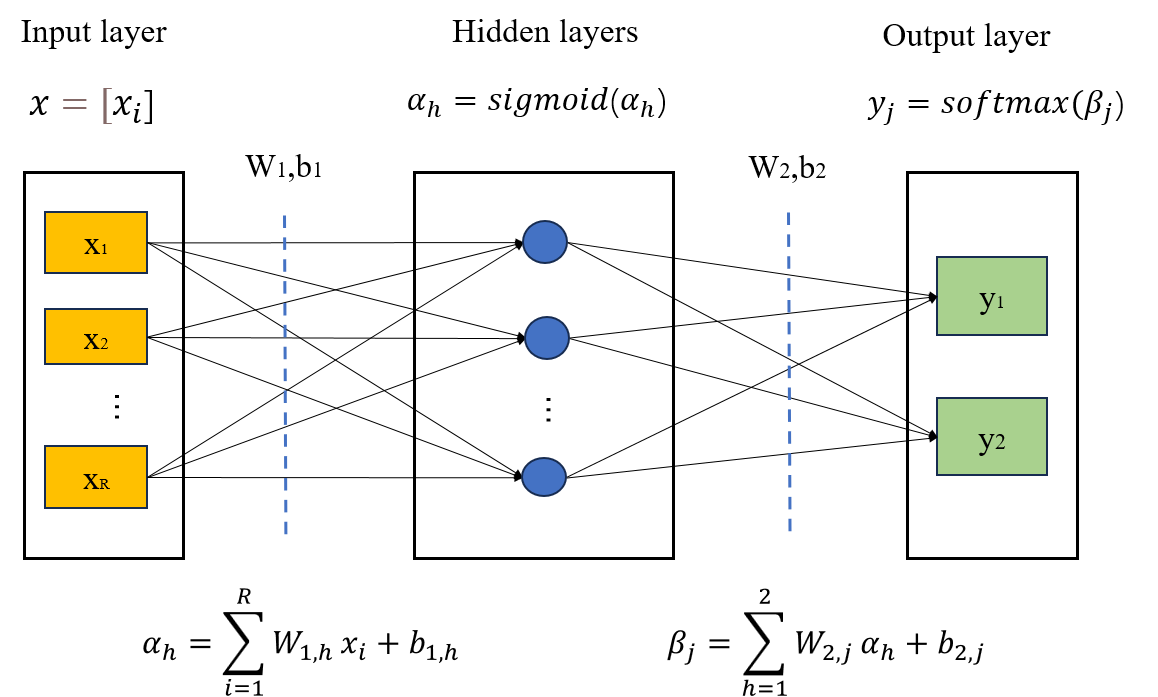}
    \caption{The structure of the neural network in this paper. The neural network is used to classify object by their extracted features.}
    \label{fig:fig6}
\end{figure*}

We design a neural network model to train by learning the mapping relationship between features in the data and their corresponding labels. We have constructed a network architecture that includes an input layer, a hidden layer, and an output layer, as shown in Fig.~\ref{fig:fig6}. The neurons in the input layer receive feature data of the input, represented by the vector $X$ = [$x_{i}$], $i$ = 1, ..., R. The number of neurons in the input layer R corresponds to the number of features in the input data. In this paper, the number of neurons in the input layer is set to R = 10, which includes ten data features: pixel count, flux, magnitude, SNR, maximum flux value, maximum background flux value, major axis, minor axis, ellipticity ratio, and inclination angle. For binary classification of stars and tracked targets, the 2 neurons in the output layer present a discrete probability distribution $Y$ = [$y_{j}$], $j$ = 1, 2, where neuron $y_{j}$ in the output layer represents the probability that the input data belongs to class $j$.


The number of neurons in the hidden layer often depends on trial and error, as it is flexible. More neurons can handle complex problems, but too few may cause underfitting, and too many can lead to overfitting. Our experiments show that at least 512 neurons are needed for good results.

The sigmoid function is employed between the input layer and the hidden layer of the model, which can improve the convergence rate of  the estimation of weights and biases. The sigmoid activation function maps variables to the range (0,1), effectively preventing overfitting and accelerating network learning. Its definition as equation~(\ref{eq:7}): 

\begin{equation}
f(x)= \frac {1}{1+e^{-x}}
\textbf{\label{eq:7}}
\end{equation}

The output layer takes the hidden layer's output as input, with the number of output nodes equal to the number of classes for classification. The activation function chosen for the output layer is the softmax function which ensures that all input values are transformed into true probabilities that sum to 1. This allows for the selection of the value with the highest probability as the predicted class. The formula used for calculating the output probabilities is equation~(\ref{eq:8}): 

\begin{equation}
g_{i}= \frac {e^{z_{i}}}{ \sum\limits_{i=1}^{2}e^{z_{i}}}
\textbf{\label{eq:8}}
\end{equation}

In the equation above, $e^{z_{i}}$ maps the classification results from $( +\infty, +\infty)$ to $(0, +\infty)$, and then normalizes them to fall between (0, 1), $g_{i}$ represents the output for learner 2 class probability values, ensuring that the sum of all output probabilities equals 1. 

Utilizing the given structure of our neural network the backpropagation algorithm is employed to refine and optimize the network's weights and biases. We choose the cross-entropy loss function \citep{de2005tutorial} to measure the distance between the actual output probability and the expected output probability. To mitigate the impact of imbalanced training sample data, we introduce weighted parameters into the cross-entropy loss function to adjust the weight ratio of different class samples (equation~(\ref{eq:9})). When considering the labels associated with the input data, a lower value of the loss function indicates a superior performance of the neural network. 
\begin{equation}
l= -{w_{i}}{ \sum\limits_{i=1}^{2}\hat{y_{i}}log({y_{i}}})
\textbf{\label{eq:9}}
\end{equation}

In this formula, $\hat{y_{i}}$ represents the actual label of the i-th sample;  $y_{i}$ represents the predicted value for the i-th sample; and $w_{i}$ represents the weight of the i-th sample.

\paragraph{Data preprocessing and equalization}~{}

\begin{figure}
\centering
	\includegraphics[width=\columnwidth]{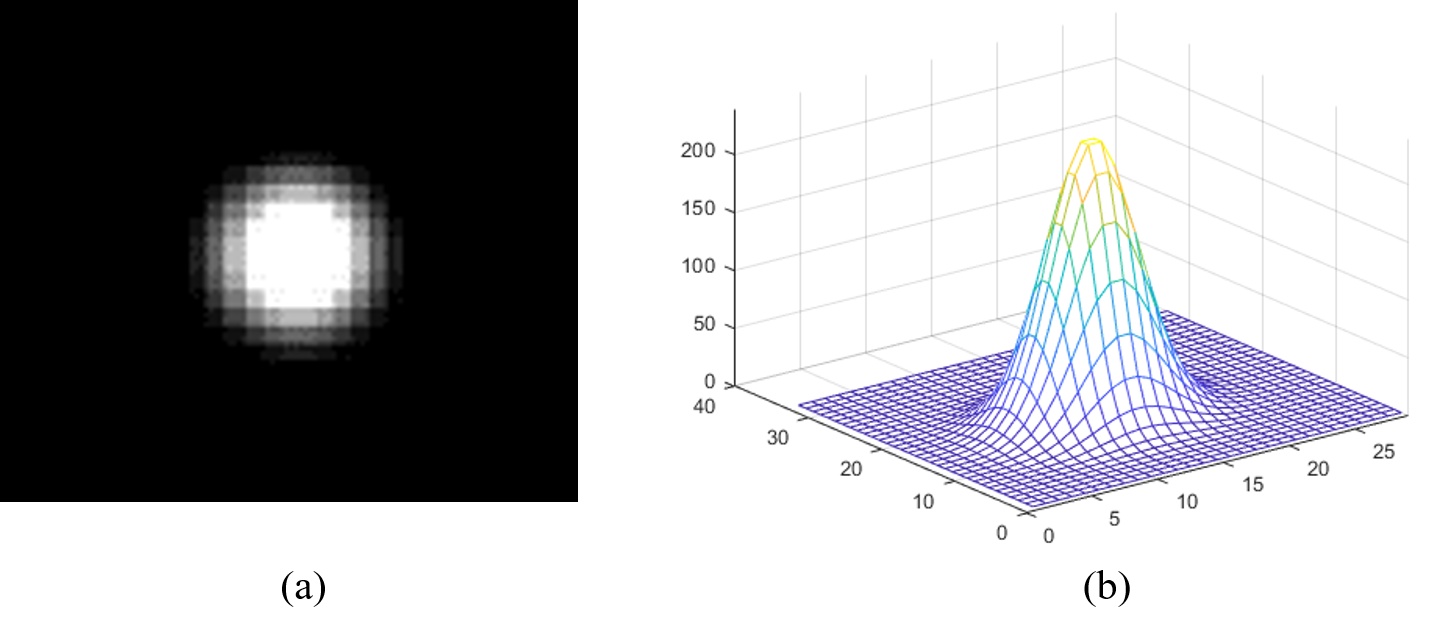}
    \caption{Simulation of tracked target image by 2D Gaussian function. (a) 2D diagram of simulated target. (b) 3D brightness distribution diagram of simulated target.}
    \label{fig:fig13}
\end{figure}

Regarding the information extracted from optical astronomical images, the data of stars are usually much more abundant than the data of tracked targets. If all the data are used for the training and testing of the classifier, the severe imbalance of the data will lead to significant deviations in the learning and testing of the classifier \citep{zhou2023machine}. To solve the classification problem of data imbalance, we use the method of extracting feature data from historical image and simulation moving object image to increase the number of moving objects and remove part of star data. We use Gaussian function as a point spread function (PSF) model to simulate the brightness distribution of star points \citep{mighell2005stellar}. In order to ensure the fidelity of the simulated image, we carefully select the standard deviation of Gaussian distribution sigma close to the real image to generate a suitable PSF, targets simulation image as shown in Fig.~\ref{fig:fig13}. The target center position and brightness are randomly distributed in the image to ensure the diversity and authenticity of the simulation data. In addition, in order to get closer to the real scene, the sky background and random noise of the real image are superimposed in the simulation image to increase the complexity of the simulation data and improve the robustness of the model to the actual noise. Due to the excessively large number of stars, we only simulated the tracked target, and all stars were based on real data.
The number of simulated tracked targets is approximately twice the amount of real tracked target data, and the number of stars is randomly reduced to no more than 100 times the target training data. This configuration of the training set helps improve model performance and mitigate biases caused by data imbalance.


Since the extracted data information has its physical meaning, we adopted the method of dividing the data by the theoretical maximum value for normalization processing. For example, the maximum values of the circularity and inclination angle of the target are 1 and $\frac {\pi}{2}$, respectively. The maximum values of the number of pixels occupied by the target, flux, instrumental magnitude, SNR, maximum flux value, maximum background flux value, major axis, and minor axis are estimated based on the actual shooting conditions. 

\paragraph{Learning procedure}~{}

PyTorch \citep{paszke2019pytorch} was used to achieve model training in this experiment. During the training process, the weights and biases were first initialized by random values and iteratively optimized through the following steps:

(1) In stochastic gradient descent, usually only one random sample (or batch) of the entire training dataset is used in each evaluation round. However, given the small size of our dataset, iterative optimization can be performed directly with the full training data. This approach improves computational efficiency and is also fully adapted to our data size. 

(2) We set the weight of equation (9) in the loss function as $w_0=1$ and $w_1$=$\frac {num_0}{num_1}$, where $num_0$ and $num_1$ represent the number of stars and targets in the training set, respectively. And adopt the gradient descent ADAM \citep{diederik2014adam} method to correct the weight and deviation, to minimize the loss function, in which the backpropagation method is adopted to achieve optimization. To control the impact of each round of global gradient estimation on the model, we set an initial learning rate of 0.001. 

(3) Repeat steps (1) - (2) to reach 10000 cycles or stop training when the relative change in loss is less than 0.001.

The parameters adopted in the specific training model are shown in Table~\ref{tab:4}.

\begin{table}
	\centering
	\caption{The neural network properties in this paper.}
	\label{tab:4}
	\begin{tabular}{cc} 
		\toprule  
    	Model property & Value\\
		\hline
        Input layer vector dimension & 12 \\
        Hidden layer size & 512 \\
        Hidden layer activation function & Sigmoid  \\
        Output layer size & 2  \\
        Output layer activation function & Softmax  \\
        Training algorithm & ADAM\\
        Learning rate & 0.001\\
	\bottomrule  
	\end{tabular}
\end{table}

\subsubsection{Further detection of moving targets}

In Section ~\ref{Neural network classifies stars and tracked targets}, we utilized stellar data as labeled data for stripe-like objects, and tracked targets data as labeled data for point-like targets. The trained neural network classification model can be applied to other image extraction data to achieve the high-precision classification effect of point-like and stripe-like targets. For point-like tracked targets, the data from a single frame image is insufficient to accurately identify which are real targets, and also to determine which belong to the same target in consecutive images. For those stripe-like targets with different moving modes, it is also necessary to distinguish them from the stripes of stars. Therefore, further detection is required to confirm the targets.

\paragraph{Using track association methods to detect tracked targets}~{}

\begin{figure}
\centering
	\includegraphics[width=.7\columnwidth]{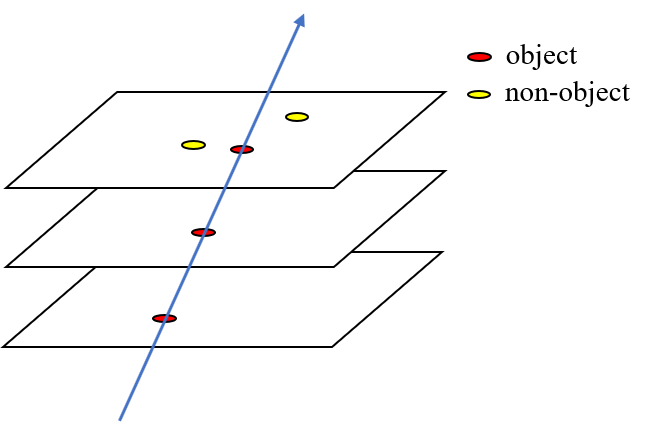}
    \caption{The method of continuous multi-frame track association is used to judge the moving object. For the image of target tracking mode, the target motion track is almost a vertical line.}
    \label{fig:fig7}
\end{figure}

For the tracked target, it appears as a point-like object, but the point-like object may also be noise, cosmic ray and others. In consecutive multi-frame images, the position, brightness, and shape of tracked targets often display a high degree of consistency and stability, whereas noise and cosmic rays appear irregularly. Therefore, after applying the neural network to classify the data, it is necessary to filter out those data. The characteristic of these noises is that the motion track in the image is inconsistent with the target, which can be filtered out by the method of motion track association, as shown in Fig.~\ref{fig:fig7}.

In this paper, the KD-Tree motion track association method used in  Section~\ref{Object detection based on clustering statistics and Track Association (TA)}  is still adopted to confirm the target. The motion track associated with at least three frames of images is determined to be the tracked target; otherwise, it is considered false alarm.

\paragraph{Using morphological statistical methods to detect targets with other motion modes}~{}

\begin{figure}
\centering  
\begin{subfigure}{\columnwidth}  
    \centering  
    \includegraphics[width=\columnwidth]{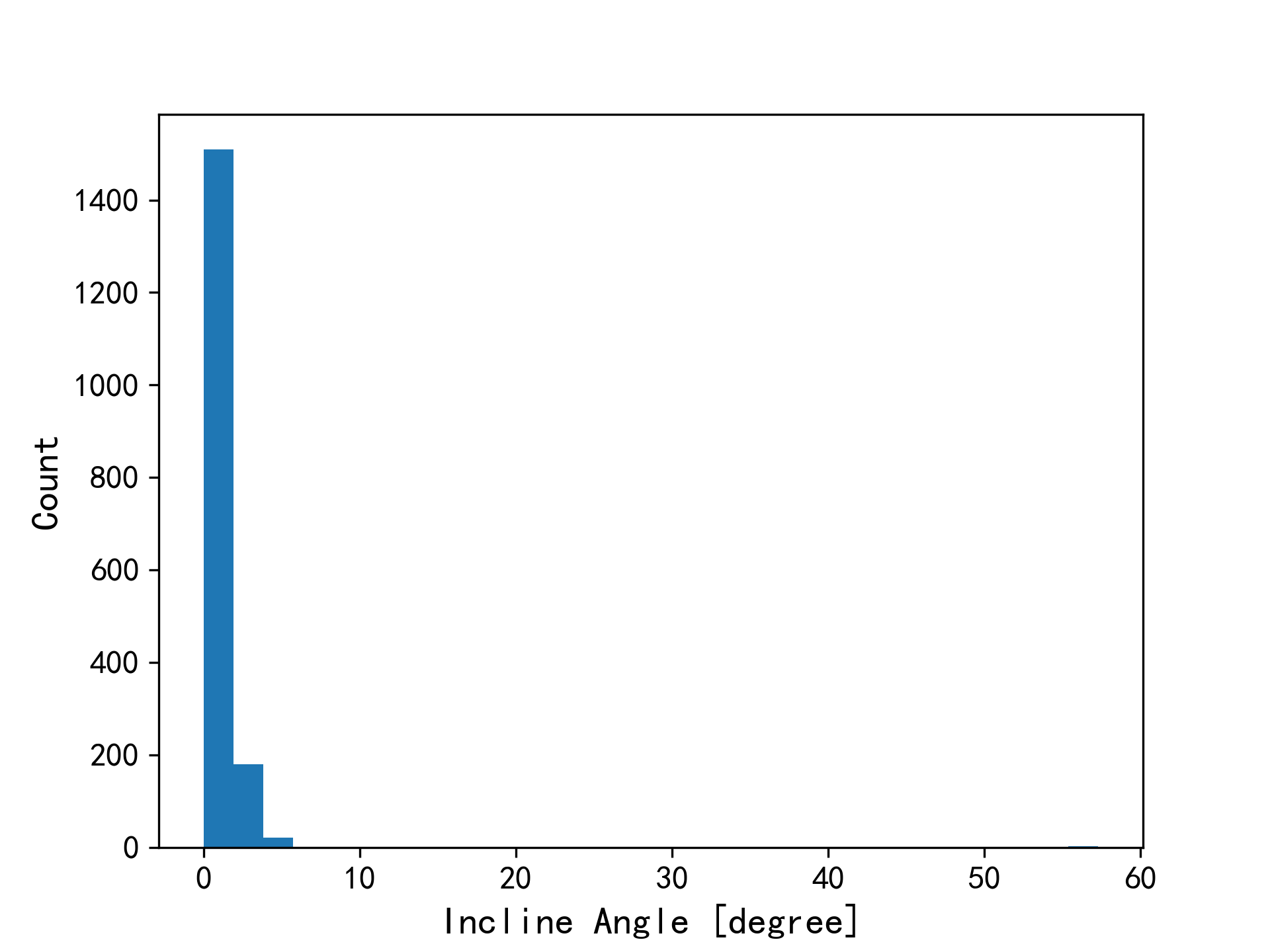}  
    \caption{}  
    \label{fig:fig8a}  
\end{subfigure}  
\hfill  
\begin{subfigure}{\columnwidth}  
    \centering  
    \includegraphics[width=\columnwidth]{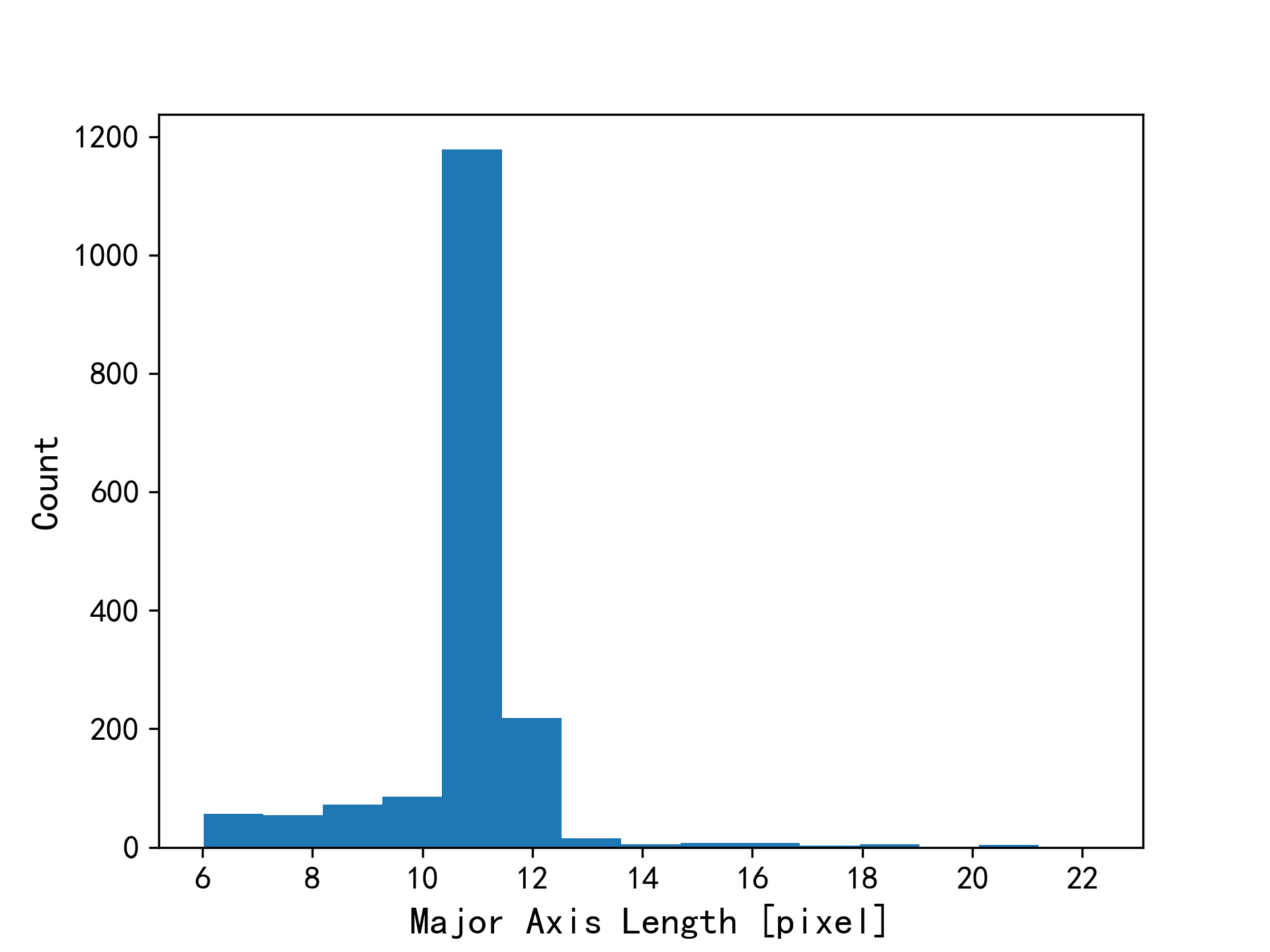} 
    \caption{}  
    \label{fig:fig8b}  
\end{subfigure}  
\caption{Histogram distribution of the inclination angle \textbf{(a)} and major axis length \textbf{(b)} of the striped target detected using the neural network model in a single frame image from Nanshan One-meter Wide-field Telescope, with a bin=30 for inclination angle and bin=15 for major axis length.}  
\label{fig:fig8}  
\end{figure}

The streak objects classified by the trained neural network include not only stars but also moving targets with other motion modes.
We can differentiate these targets from stars by analyzing the inclination and length of the streaks. When tracking targets, due to the relative motion of the star and the tracked target, stars will appear as straight streaks in a specific direction on the image plane. In this study, we conducted a statistical analysis on two features: the inclination angle and major axis length. Fig.~\ref{fig:fig8} illustrates the histogram distribution of these features for streaks  identified by the neural network. The analysis indicates that the feature set with the highest proportion in the data represents stars, while other distributions likely correspond to moving objects under different motion modes.

For the extracted inclination angle data in figure~\ref{fig:fig_angle}, we set the representation range of the inclination angle to ±90°.
The extraction of the major axis lengths of star streaks, particularly for faint stars with a low SNR, can be challenging. These dim streak signals may be partially obscured by background noise, leading to inconsistencies in the extracted major axis length information compared to those derived from brighter and clearer stellar streaks.

In order to evaluate this property, we use the method of data distribution statistics to estimate the possible distribution range of star stripes through the statistics of their inclination angle and major axis length. For the distribution of data in the maximum interval (a, b),
we set the number of bars in a histogram as 'bin' and divide the data into 'bin' equal-width intervals. For each such interval, we count the number of parameter values that fall within it, denoted as $n_k$, where k represents the interval index ($k$ = 1, 2, ..., bin). 
At the same time, the interval with the most concentrated parameter values is found, that is, the interval corresponding to the maximum frequency is found and the mean of the interval is calculated, where the inclination angle is represented as $mean_{\theta}$ and the length of the major axis is represented as $mean_{L}$. In this paper, moving objects search rules for other motion modes are as follows:

(1) The objects classified as star by the neural network model are screened, and the objects with too short major axis and too small pixel area are removed.

(2) For the inclination angle distribution, it is considered that the moving object is when the difference between the inclination angle and $mean_{\theta}$ is greater than 10°.

(3) Targets whose difference between the inclination angle and $mean_{\theta}$ is less than 10° and whose difference between the major axis length and $mean_{L}$ is more than 15 pixels are also considered as moving objects (The parameters are set based on multiple experiments).

(4) Because sources with too large connected area greater than 600 pixels are filtered out in the detection stage, the method does not detect targets with fast moving speed in the image, that is, excessively long streaks.

\section{ Results } \label{sec:style}
\subsection{ Experimental setting} \label{subsec:tables}

In this section, we use actual observation images to verify the performance of the moving objects detection method. The measured images used in the experiment come from Nanshan One-meter Wide-field Telescope \citep{bai2020wide} with a diameter of 1m, a focal length of 2200 mm, and a field radius of 1.5°×1.5°. The CCD camera was designed and integrated by the CCD laboratory of the National Astronomical Observatories of China (NAOC) with a 4096 × 4136 imaging pixels, the relevant parameters of the telescope and its detectors are shown in the Table~\ref{tab:telescope}.

\begin{figure}
\centering
	\includegraphics[width=\columnwidth]{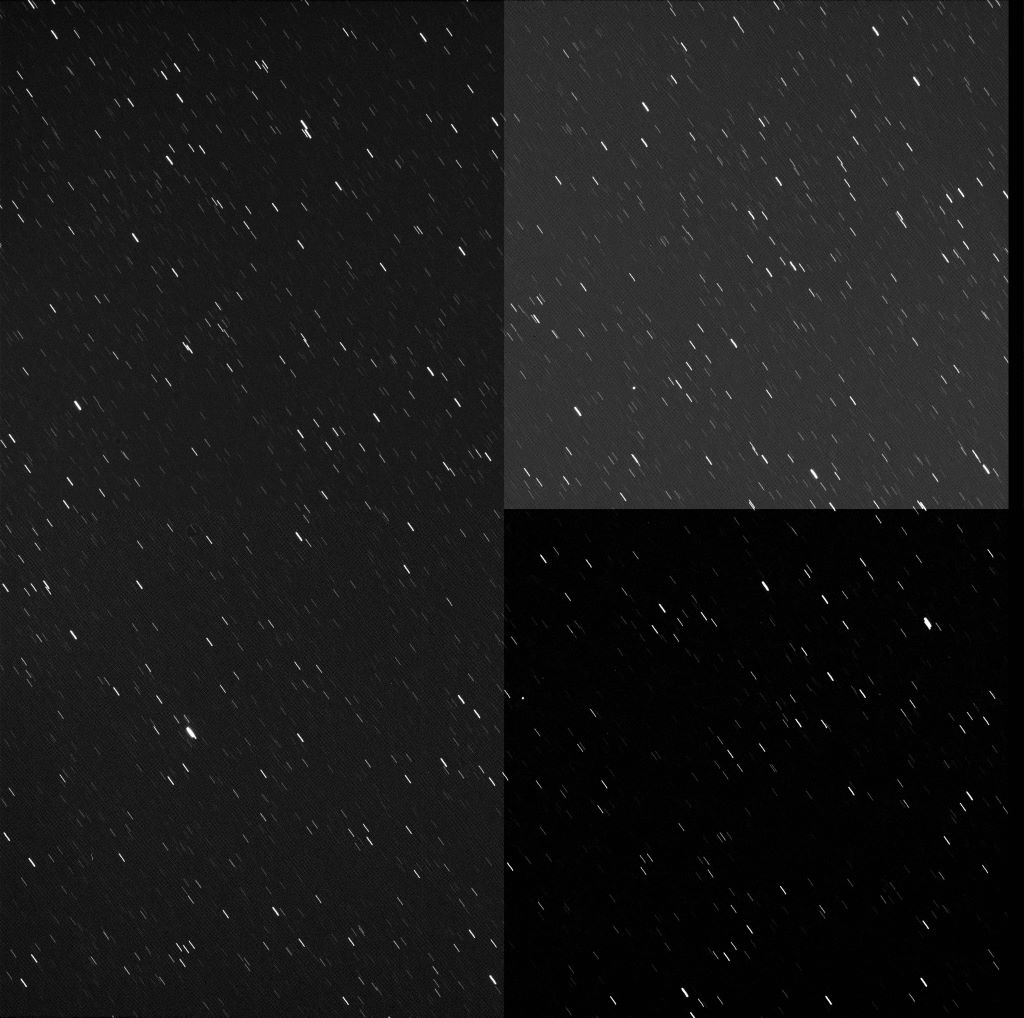}
    \caption{An image of space objects captured by Nanshan One-meter Wide-field Telescope at the Nanshan Observatory of the Xinjiang Astronomical Observatory in June 2018. The image has not undergone bias and flat field correction.}
    \label{fig:fig9}
\end{figure}

Experimental images were taken in June 2018, during a moonless night with good weather conditions at the Nanshan Observatory of the Xinjiang Astronomical Observatory, the band was captured in the R-band of Johnson-Cousins UBVRI system \citep{Bessell_1990}. Under the target tracking mode with an exposure time of 3s and the tracking speed of the telescope is approximately 21 ''/s, stars appear as streaks, while tracked targets appear as points. We have selected 43 different sky regions for observation, capturing 3 frames of images for each region, totaling 129 frames of images. And continuously tracking the two targets to obtain 62 frames and 21 frames of images respectively. So we have a total of 212 frames of FITS images for the experiment, one of the images is shown shown in Fig.~\ref{fig:fig9}. We selected the data from the first 99 frames of 129 images in chronological order as the real data in the training set and the data from the last 30 frames of 129 images in chronological order as the test dataset. Additionally, we combined these 30 frames of test data with continuous tracking of 62 frames and 21 frames of images to form the validation dataset. The number of real tracked targets in experimental images is shown in the Table~\ref{tab:num}.

In this paper, the experiments were conducted using c/c++ as the Feature Extraction Software development environment and Python 3.10 as the primary programming language for the implementation of the proposed method. The hardware configuration for the experiments includes an Intel 8-core CPU with 3GB of RAM, coupled with an NVIDIA GeForce GT 710 graphics card.

\begin{table}
	\centering
	\caption{The relevant parameters of Nanshan One-meter Wide-field Telescope and its detectors.}
	\label{tab:telescope}
	\begin{tabular}{cc} 
		\toprule  
    	Features &  Characteristics \\
		\hline
        Effective diameter & 1000 mm \\
        Prime focal length & 2200 mm \\
        FOV at prime focus & 1.5°×1.5° \\
        Primary mirror reflectance & 87\%  \\
        Primary mirror material & Schott Zerodur  \\
        CCD pixel number & 4096 × 4136\\
        CCD pixel size & 12 $\mu$ m $ \times$ 12 $\mu$ m\\
        CCD pixel scale & $1.125^{''}$\\
         Dark current ($e^{-}$ pixel$^{-1}$ h$^{-1}$) & 3 $@$ 300 K (i.e., $-100^\circ$ C) to 293 K (i.e., $0^\circ$ C), 0.01 $@$ 300 K to 120 K (i.e., $-153^\circ$ C) \\  
        Readout noise & $3 \sim 4.5 e^{-}$ \\
        Full frame readout time & 27 s $@$ 146 kHz, 44 s $@$ 91 kHz, 78 s $@$ 51 kHz \\         
        Linearity & > 99.9995 $\% $\\
	\bottomrule  
	\end{tabular}
\end{table}

\begin{table}
	\centering
	\caption{The number of real tracked targets in experimental images, where the count is determined manually. The 129 images in the table refer to 129 frames from 43 different sky regions for observation.}
	\label{tab:num}
	\begin{tabular}{cc} 
		\toprule  
    	Image source &  Number of real tracked targets \\
		\hline
        The first 99 frames of 129 images & 360 \\
        The last 30 frames of 129 images & 141 \\
        62 frames of continuous observation & 62 \\
        21 frames of continuous observation & 42  \\
	\bottomrule  
	\end{tabular}
\end{table}


\subsection{Evaluation metrics} \label{subsec:tables}

To evaluate the performance of moving object detection methods effectively, we take the moving objects determined by manual judgment as the ground truth. We utilize confusion matrix, detection probability, and false alarm rate as key metrics. The number of detected moving objects is set as TP, the number of undetected moving objects is FN, and the number of false targets detected as  moving objects is FP. The detection probability ($P_d$) refers to the probability that a method correctly detects targets, which is represented by the proportion of correctly detected targets to the actual total targets in this paper. And false alarm rate ($FAR$) refers to the probability that a method erroneously identifies a non-target signal as a target. In this paper, it is represented by the ratio of the number of non-targets detected as targets to the sum of the number of actual targets and the number of non-targets detected as targets \citep{liu2020space}.
$P_d$ and $FAR$ are defined as equation~(\ref{eq:10}) and (\ref{eq:11}):

\begin{equation}
P_d = \frac{TP}{TP + FN}  
\textbf{\label{eq:10}}
\end{equation}

\begin{equation}
FAR = \frac{FP}{ FP+FN+TP }  
\textbf{\label{eq:11}}
\end{equation}

\subsection{Experiment} \label{subsec:tables}

\subsubsection{Feature extraction based on the Feature Extraction Software}

For these 212 frames of FITS images used in the experiment, we use the Feature Extraction Software to achieve source's feature extraction from images, and the feature information of all sources whose threshold value is greater than equation~(\ref{eq:2}) is obtained. Table~\ref{tab:2} shows several groups of feature data randomly selected  by us.

\begin{table*}
	\centering
	\caption{Partial feature data extracted from the Feature Extraction Software.}
	\label{tab:2}
	\begin{tabular}{lcccccccccccccr} 
		\toprule  
		Frame & Img-x & Img-y & Pixels-N &	Flux &INST-mag&	SNR	& Pix-flux-peak &Pix-flux-bk & Major & Minor & Circularity & Incline\\
		\hline

1&	2092.4	&2781.6&128&	170067.0	&13.1&	403.7&	16304.0&	57.8	&1.6 &	1.5&	0.96&	0.75\\
55&	2570.0&	3361.0	&89	&62475.9&	14.2&	239.1&	4864.6	&65.0&	1.6&	1.6&	0.99&	0.88\\
16 &	823.8&	602.3&	322	&571501.3&	11.8&	741.6&	43405.3&	69.5&	1.9&	1.7&	0.89&	0.87\\
13&	2864.5&	1423.3&	6546&	23641728&	7.8&	4784.2&	54988.1&	118.8&	20.1&	5.7	&0.28	&1.40\\

17&	1146.0&	4129.4&	2369&	138600.8&	13.3&	302.3&	510.5&	30.2&	96.0&	4.2&	0.04&	0.09\\
58&		4091.3&	3023.1 &	13&	617.8&	19.2&	18.0&	84.4&	43.2&	1.3&	1.3&	0.25&	0.79\\
9 &	2366.5&	242.5 &	1583 &	9797929 &	8.7 &	3111.7 &	55034.2 &	73.9 &	11.6 &	2.0 &	0.17 &	1.11\\
1&	2502.8&	3281.5	&91&	63821.1&	14.2&	240.3&	5281.9&	73.9&	1.6&	1.6&	0.40	&0.88\\
2&	2128.3&2867.6	&971&	2681195.8&	10.1	&1619.7&	21457.9&	60.9&	12.0&	2.0&	0.17&	1.12\\
81	&	1555.4	&3503.9&	281	&855806.3&	11.4&	908.3&	55004.8&	113.6&	1.8&	1.8&	0.98&	0.76\\

... & ... & ... & ... & ... & ... & ...& ...& ...& ...& ...& ...& ... \\

		\bottomrule  
	\end{tabular}
\end{table*}

\subsubsection{Automatic labeling of stars and tracked targets}

Because we select the data of the first 99 frames in time order from 129 images as the real data in the training set, we need to automatically label these 99 frames of images.

\begin{figure*}
\centering
	\includegraphics[width=0.8\textwidth]{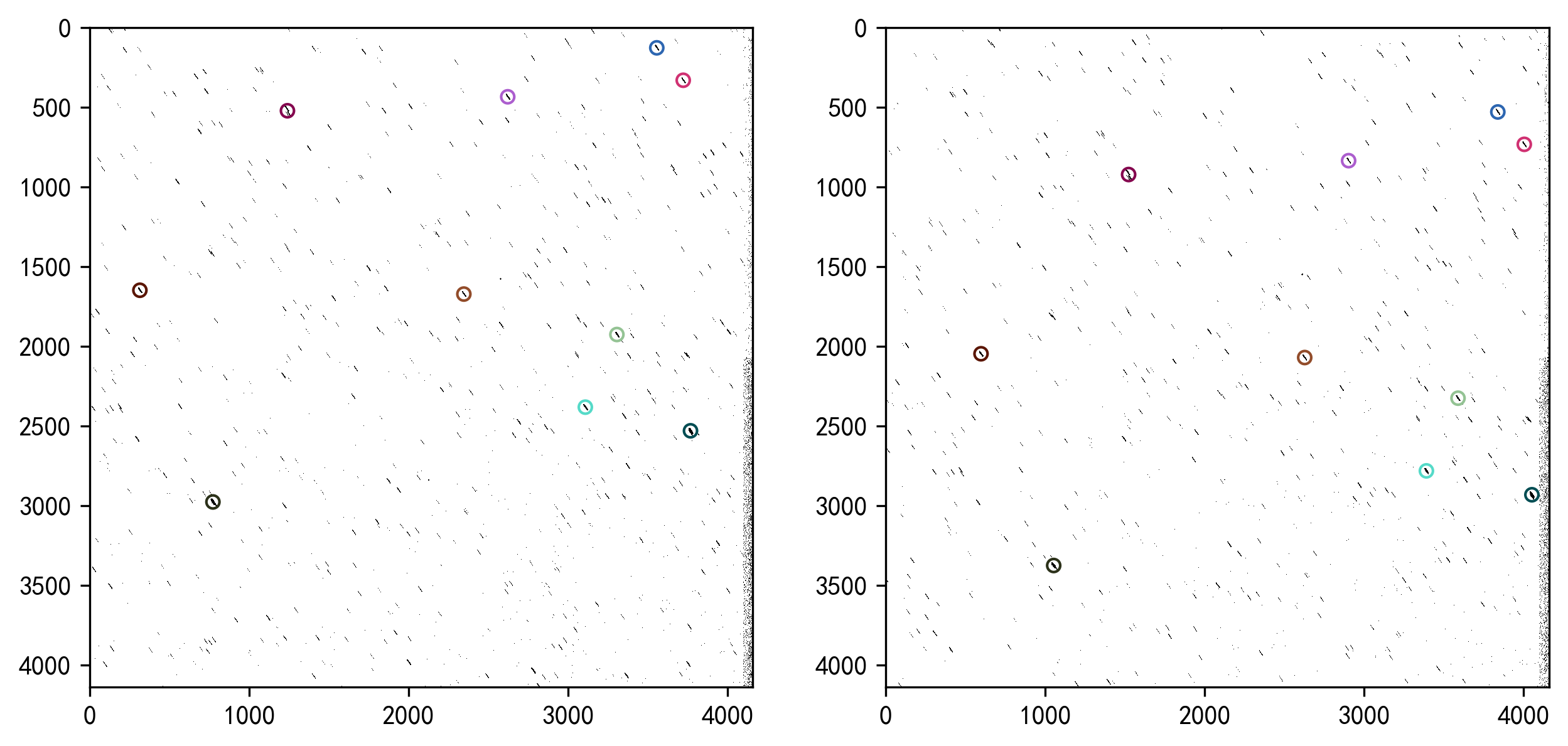}
    \caption{Star align based on image registration. Left and right are two consecutive frames of observation images, and the circles in the image mark the 10 brightest stars in terms of flux. The circle of the same color in the left and right images represents the same star matched. We performed a black-and-white color display and inversion on the image.}
    \label{fig:fig10}
\end{figure*}

\begin{figure}
\centering  
\begin{subfigure}{\columnwidth}  
    \centering  
    \includegraphics[width=\columnwidth]{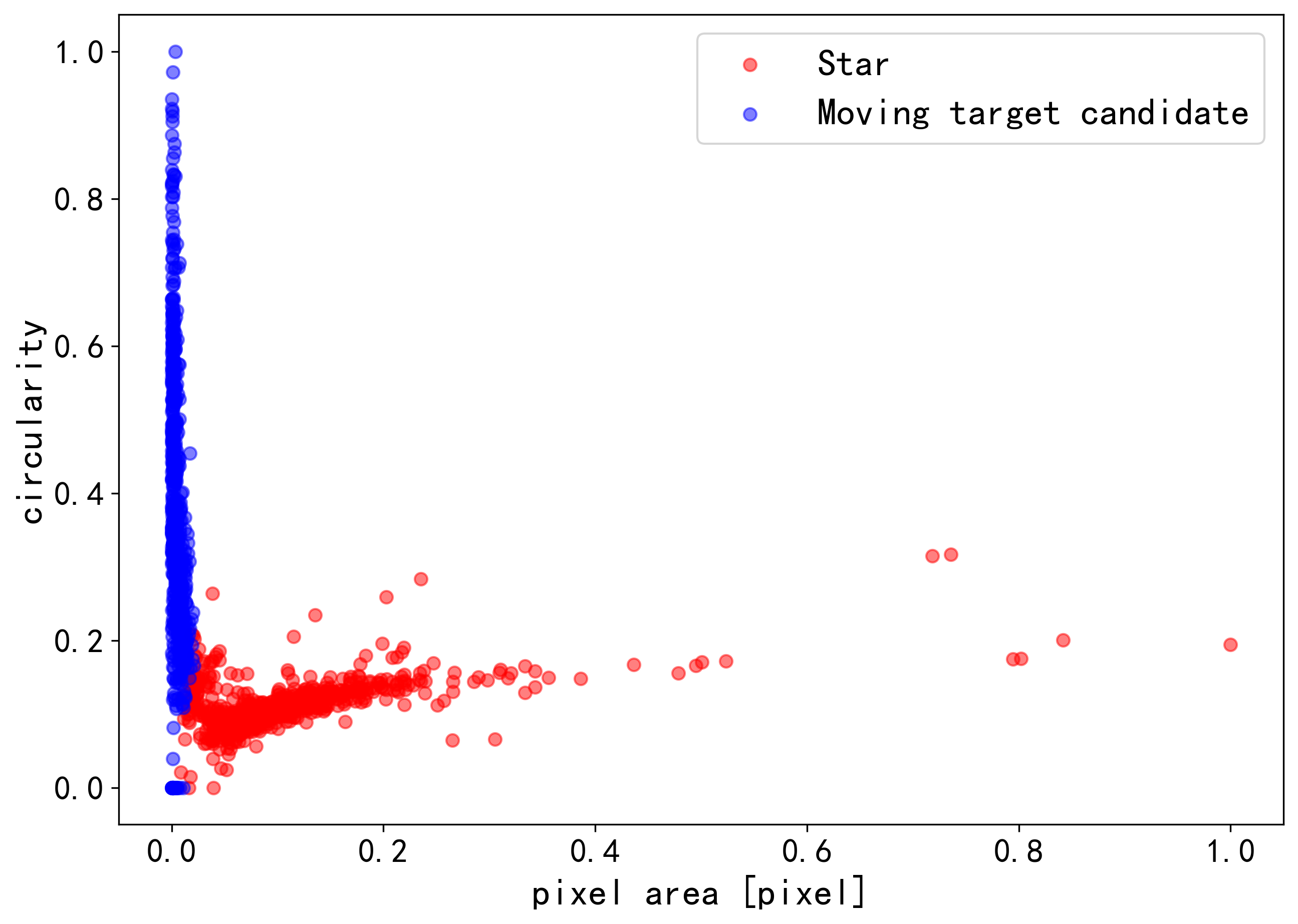}  
    \caption{}  
    \label{fig:fig11a}  
\end{subfigure}  
\hfill  
\begin{subfigure}{\columnwidth}  
    \centering  
    \includegraphics[width=\columnwidth]{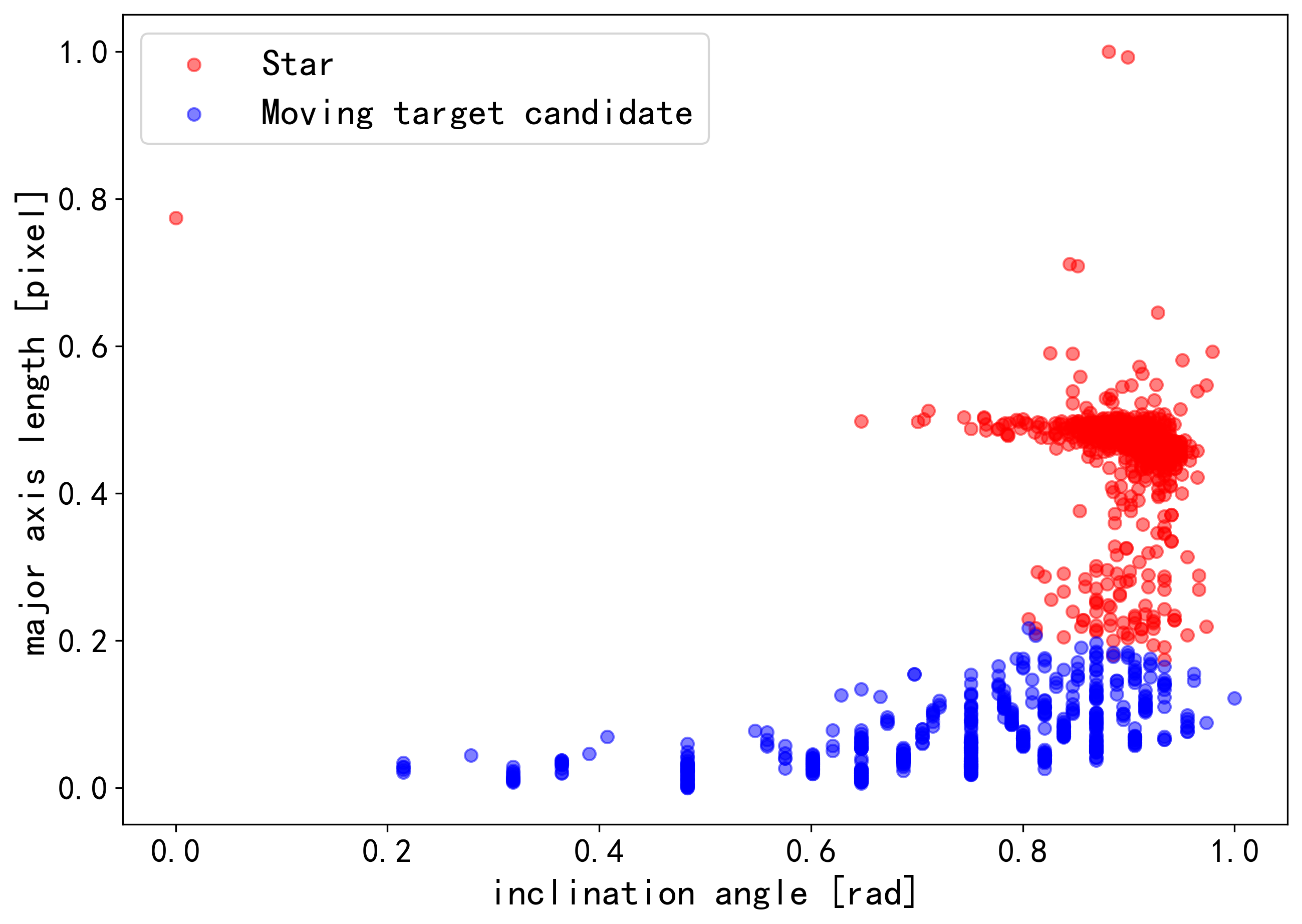} 
    \caption{}  
    \label{fig:fig11b}  
\end{subfigure}  
\caption{Star and moving object candidate data classified by KMeans method in three consecutive frames. \textbf{(a)} shows the classification results of pixel area and circularity. \textbf{(b)} shows the classification results of inclination angle and major axis length. The data is regularized.}
\label{fig:fig11}  
\end{figure}

Based on the image registration method, stars aligned in two consecutive frames of images can be obtained. The result example is shown in Fig.~\ref{fig:fig10}. The two images in the figure are two frames of images taken continuously, in which the circles in the image mark the 10 brightest stars in terms of flux, and the stars of the same color in the two frames are matched stars. The star data detected after image registration is removed and the target candidate data is obtained by k-means classification. The classification results are shown in Fig.~\ref{fig:fig11}, where Figure \ref{fig:fig11} (a) shows the classification results of pixel area and circularity, while Figure \ref{fig:fig11} (b) displays the classification results of inclination angle and major axis length. It can be observed from the figures that this method can effectively separate stars from candidate targets.The tracked target data is obtained via the track association method for the tracked targets candidate.

We have analyzed the results of the automatic annotation. Through manual detection, the actual number of tracked targets in 99 frames of images is 360, while the number of automatically annotated targets is 260, and there are no false alarm targets. The number of labeled stars is 106,011.
We label the identified star data as 0 and the identified tracked target data as 1. It should be noted that in the process of target detection, a single star may be identified as multiple stars due to the discontinuity of a small number of faint star streaks, and there are also cases where blended stars are identified as the same star, and these special star data are still used as the feature data of stars.

\subsubsection{Classification of stars and tracked targets based on neural network method}

Due to the relatively small amount of data for the targets compared to the stars, we achieved the expansion of target data by simulating target images and extracting data. A total of 24 frames of images were simulated, with 23 targets in each frame, totaling 552 tracked targets. We used Feature Extraction Software to process the simulated images to obtain more parameters of moving objects, while randomly reducing the data quantity of stars by 40\%. Therefore, the total number of tracked targets in the training dataset is 812 and the number of stars is 63,607. 
We choice the data from the last 30 frames of 129 images in chronological order as the test dataset. Additionally, we combined these 30 frames of test data with continuous tracking of 62 frames and 21 frames of images  to form the validation dataset. the number of tracked targets and star in the data set is shown in the Table~\ref{tab:num1}.

\begin{table}
	\centering
	\caption{The number of stars and tracked targets in different datasets, where the number of stars in the test dataset and validation dataset includes unknown noisy data.}
	\label{tab:num1}
	\begin{tabular}{ccc} 
		\toprule  
    	Data set &  Number of tracked targets  &  Number of star \\
		\hline
        training dataset & 812 & 63,607 \\
        test dataset & 141 & 66,491 \\
        validation dataset & 245 & 283,727 \\

	\bottomrule  
	\end{tabular}
\end{table}

\begin{figure}
\centering  
\begin{subfigure}{0.8\columnwidth}  
    \centering  
    \includegraphics[width=\columnwidth]{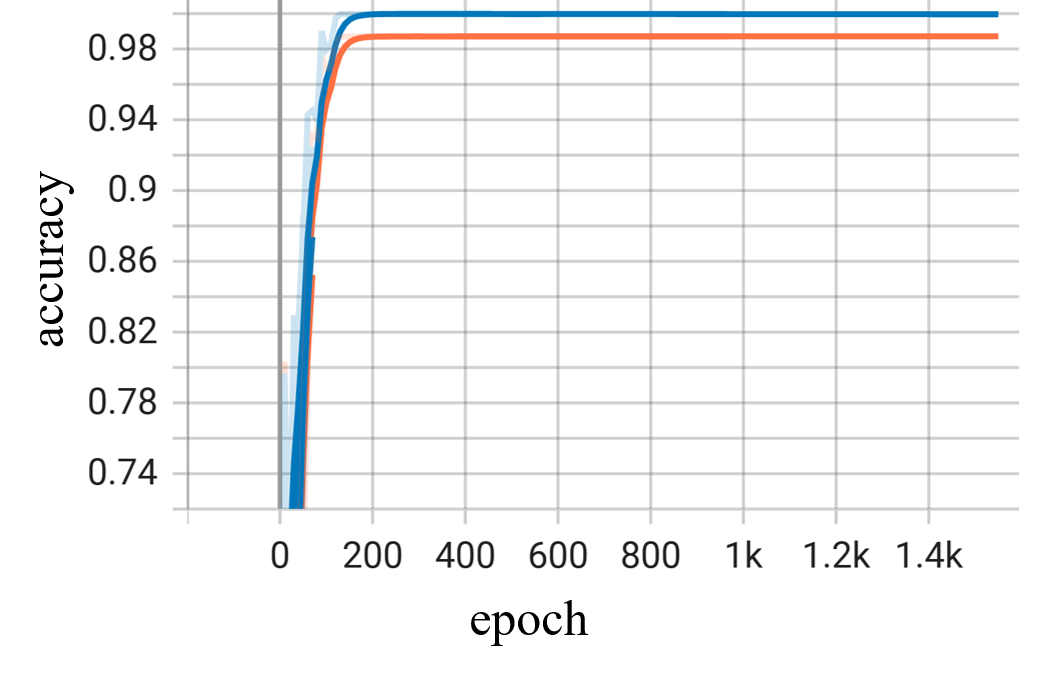}  
    \caption{}  
    \label{fig:fig11a}  
\end{subfigure}  
\hfill  
\begin{subfigure}{0.8\columnwidth}  
    \centering  
    \includegraphics[width=\columnwidth]{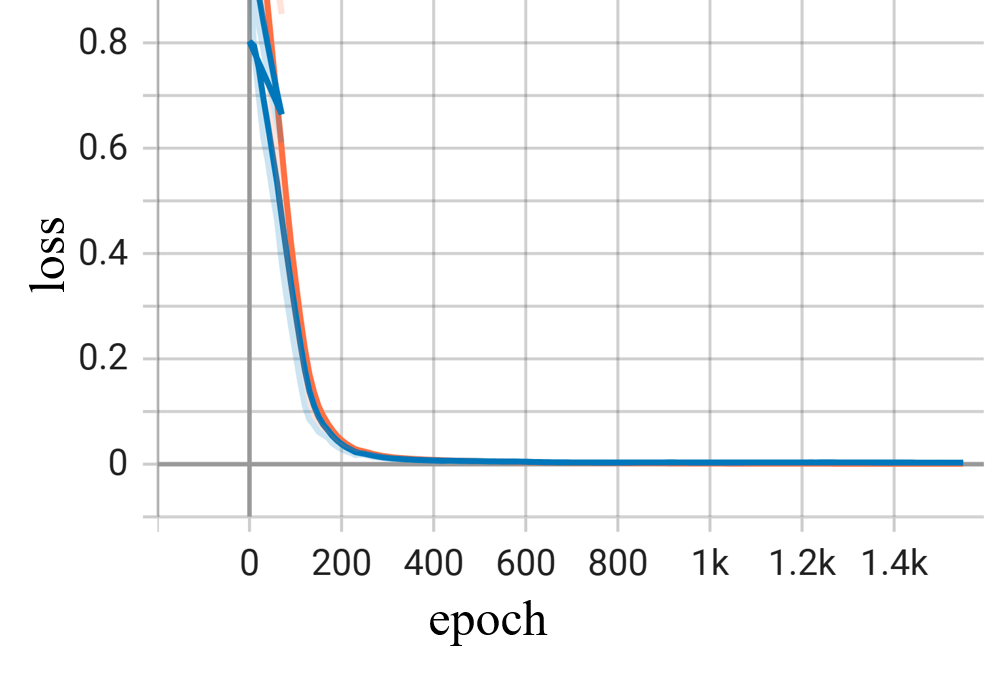} 
    \caption{}  
    \label{fig:fig11b}  
\end{subfigure}  
\caption{The accuracy and loss of the model classification vary with epochs, the blue line represents the training set and the red line represents the test data set.}
\label{fig:fig14}  
\end{figure}

We used the neural network model constructed in Section ~\ref{Neural network classifies stars and tracked targets} to classify and test the training set and test set. We set the weight of equation (9) in the loss function as $w_0=1$ and $w_1=63607/812$, and obtained the corresponding training results.
The accuracy and loss values changes in the model training with epoch are shown in Fig.~\ref{fig:fig14}. It can be seen from the images that the training effect of the model is good without overfitting phenomenon.

\begin{figure}
\centering  
\begin{subfigure}{\columnwidth}  
    \centering  
    \includegraphics[width=\columnwidth]{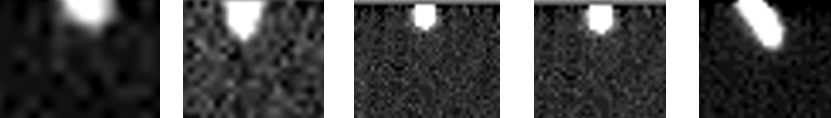}  
    \caption{Cropped star}  
    \label{fig:fig15a}  
\end{subfigure}  
\hfill  
\begin{subfigure}{\columnwidth}  
    \centering  
    \includegraphics[width=\columnwidth]{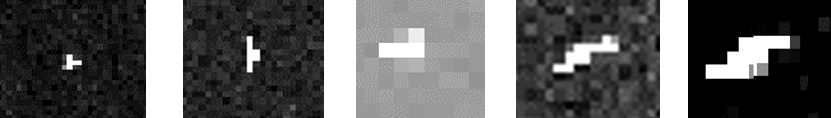} 
    \caption{Cosmic rays}  
    \label{fig:fig15b}  
\end{subfigure}  
\hfill  
\begin{subfigure}{\columnwidth}  
    \centering  
    \includegraphics[width=\columnwidth]{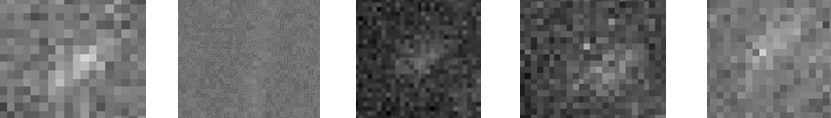} 
    \caption{Faint star streaks or noise}  
    \label{fig:fig15b}  
\end{subfigure}  

\caption{The image shows the types of point-like targets detected by the neural network model in the validation dataset, excluding the tracked target.}  
\label{fig:fig15}  
\end{figure}

During the data inspection process, we found that in addition to the tracked targets detected as point-like data, there are also the following types of data:

(1) Cropped star: as shown in Fig.~\ref{fig:fig15} (a) these are stars that appear on the edge of the frame. The imaging of stars should normally be streak-shaped, but due to the limitations of the field of view, Cropped stars appear on the edges of the image, which can easily be mistaken for tracked objects. 

(2) Cosmic rays: as shown in Fig.~\ref{fig:fig15} (b), cosmic ray noise consists of jets formed by various high-energy particles from space, primarily including protons and various other particles. The brightness value of cosmic ray noise is significantly higher than that of the surrounding pixels and is distributed randomly. Since the training data only includes stars and targets, cosmic rays may be mistakenly identified as tracked objects. 

(3) Faint star streaks or noise: as shown in Fig.~\ref{fig:fig15} (c), when a star streak is too faint, it can be difficult to match in the star matching process, and the results of feature extraction such as centroid position have considerable errors, which may lead to its false detection as tracked objects.

The tracked target in the image cannot be confirmed only from the morphological information of the target in a single frame image. We also need to associate the moving track of the target with the time information to confirm whether the changed target is the tracked target.

\subsubsection{Further detection of moving targets}

(1) Track association for detection of the tracked target

\begin{figure}
\centering
	\includegraphics[width=\columnwidth]{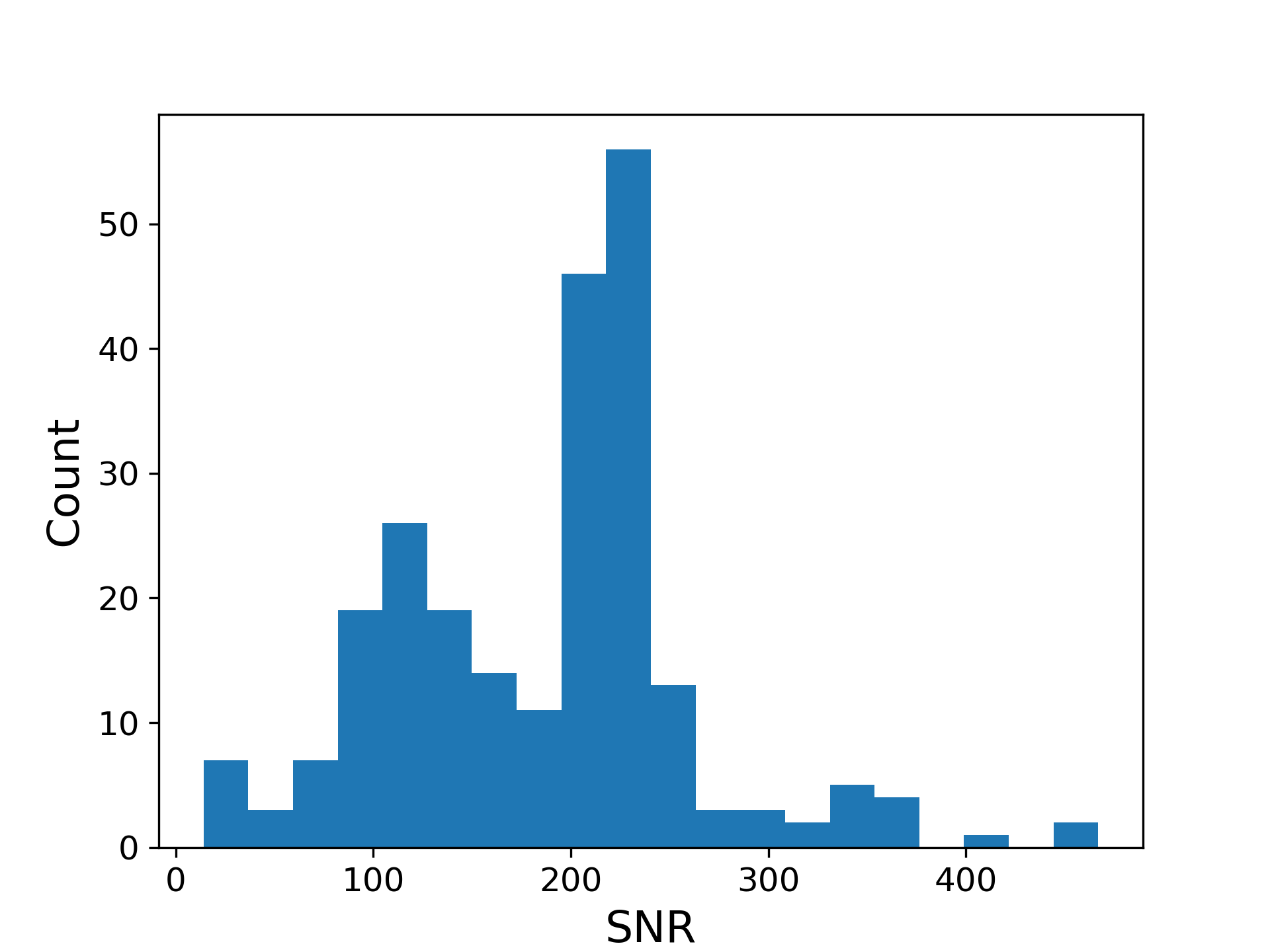}
    \caption{SNR distribution histogram of 245 real tracked targets in the validation dataset.}
    \label{fig:fig16}
\end{figure}

The number of real tracked targets in the validation dataset is 245. After performing track association with a threshold of R=30, the detection results for these tracked targets are presented in Table~\ref{tab:5}. The SNR distribution of these 245 real tracked targets is shown in Fig.~\ref{fig:fig16}.



\begin{table}  
\renewcommand{\arraystretch}{1}  
\caption{Detection results of tracked targets (TT) in validation dataset.}  
\label{tab:5}  
\centering  
\begin{tabular}{cccc}  
\toprule  
& &\multicolumn{2}{c}{Detection} \\  
& Class & TT & Other \\  
\midrule  
\multirow{2}{*}{Ground Truth}   
& TT & 239 & 6 \\  
& Other & 0 & - \\  
\midrule  
\multicolumn{2}{c}{$P_d$} & \multicolumn{2}{c}{97.55$\%$} \\  
\multicolumn{2}{c}{FAR} & \multicolumn{2}{c}{0$\%$} \\  
\bottomrule  
\end{tabular}  
\end{table}

\begin{figure}
\centering  
\begin{subfigure}{0.45\columnwidth}  
    \centering  
    \includegraphics[width=\columnwidth]{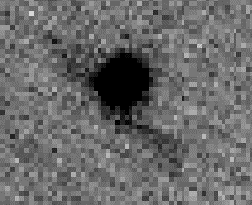}  
    \caption{}  
    \label{fig:fig11a}  
\end{subfigure}  
\hfill  
\begin{subfigure}{0.45\columnwidth}  
    \centering  
    \includegraphics[width=\columnwidth]{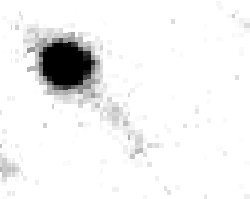} 
    \caption{}  
    \label{fig:fig11b}  
\end{subfigure}  
\caption{Images of star overlapping with object. In this case, it is difficult for the neural network model to detect the targets. }  
\label{fig:fig17}  
\end{figure}

The analysis of missed targets found that it was difficult to detect such targets when the moving object overlapped with the star image, as shown in Fig.~\ref{fig:fig17}. At the same time, it was found that if the number of consecutive image frames was small, part of the tracked target would not be associated, and the target would be omitted.

The method proposed in this paper uses the neural network model to classify objects of different shapes in a single frame image, and the track association method is used to remove false alarms from the model classification results. The automatic naive labeling method in this paper also adopts such a process. Therefore, this paper applies the naive target classification method to the validation dataset, and the threshold value R of track association is also set to 30. The detection results of the two methods are shown in Table~\ref{tab:6}.

\begin{table}
	\centering
	\caption{Under the same environment with an Intel 8-core CPU and 3GB of RAM, this table compares the performance of the proposed target tracking method with the naive detection method on the validation dataset using the Python 3.10, where the time refers to the average processing time of a single frame image.}
	\label{tab:6}
	\begin{tabular}{cccc} 
	\toprule  
    	method & $P_d$ & FAR &  time\\
		\hline
		This paper & 97.55 $\%$ & 0 & 0.33s\\
        A naive method & 73.47 $\%$ & 23.67 $\%$ & 7.01s\\
	\bottomrule  
	\end{tabular}
\end{table}

The naive target classification takes a long time for star image registration calculations, and using the same parameters in the k-means clustering method after removing stars does not classify well in some images. Therefore, compared to the neural network model, its accuracy and efficiency are poorer. Directly using a neural network model can reduce the error rate of target detection due to manual threshold settings, and it is equally suitable for real-time target detection as well.

(2) Morphological statistical methods for the detection of targets with other motion modes.

After manual verification, the validation set contains a total of 39 targets with other motion modes, and the detection results for these targets are shown in Table~\ref{tab:7}. In the experiment, it was found that the inclination and major axis length of overexposed stars can change, leading to false alarms generated by this method for overexposed stars.

\begin{table}  
\renewcommand{\arraystretch}{1}  
\caption{Detection results of targets with other motion modes (TOM) in validation dataset.}  
\label{tab:7}  
\centering  
\begin{tabular}{cccc}  
\toprule  
& &\multicolumn{2}{c}{Detection} \\  
& Class & TOM & Other \\  
\midrule  
\multirow{2}{*}{Ground Truth}   
 &TOM    & 30 & 9  \\
&Other  & 15 & -  \\
\midrule  
\multicolumn{2}{c}{$P_d$} & \multicolumn{2}{c}{76.92 $\%$} \\  
\multicolumn{2}{c}{FAR} & \multicolumn{2}{c}{27.78 $\%$} \\  
\bottomrule  
\end{tabular}  
\end{table}  

\subsection{Results and analysis} \label{subsec:tables}
\begin{figure*}
\centering  
\begin{subfigure}{0.9\columnwidth}  
    \centering  
    \includegraphics[width=\columnwidth]{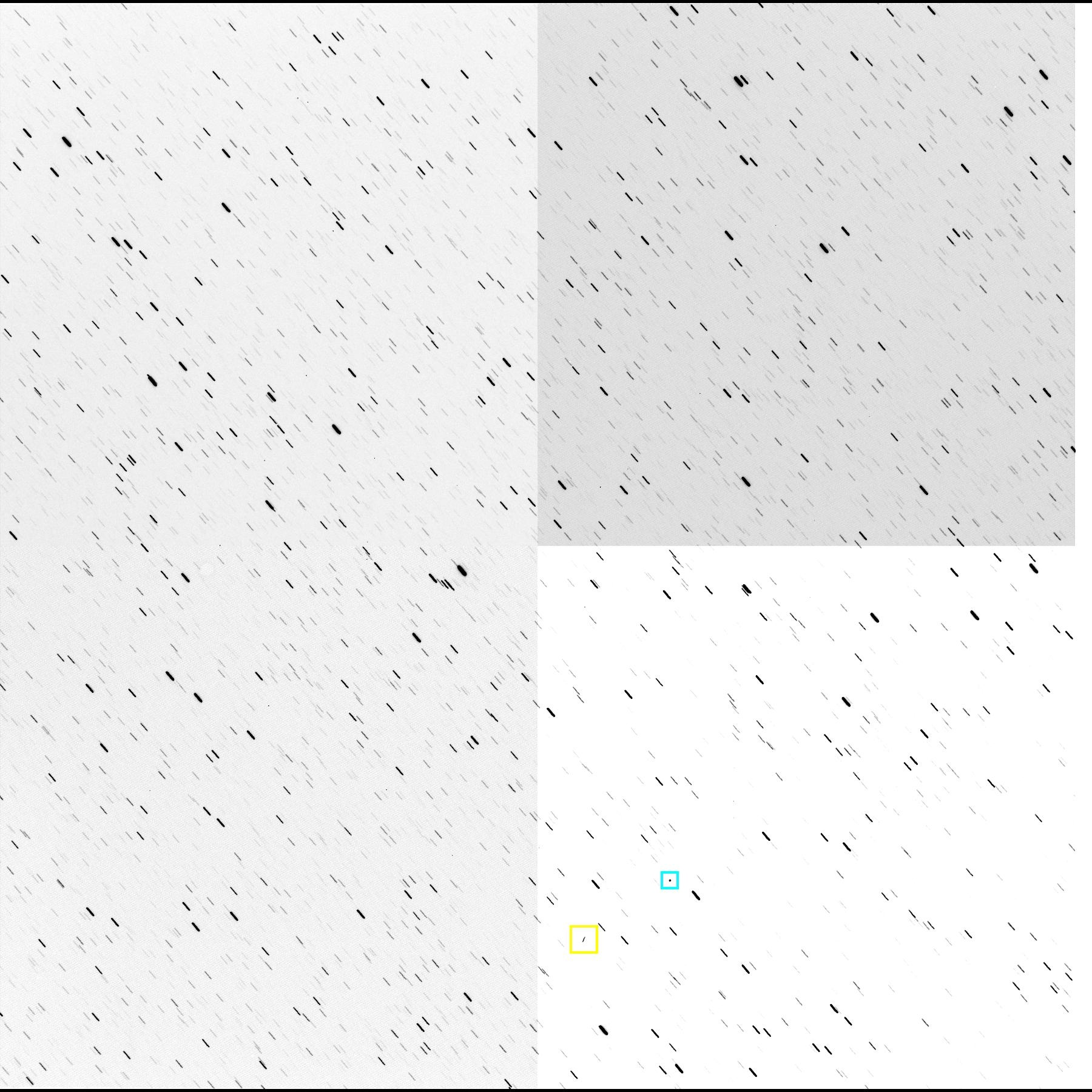}  
    \caption{}  
    \label{fig:fig11a}  
\end{subfigure}  
\begin{subfigure}{0.9\columnwidth}  
    \centering  
    \includegraphics[width=\columnwidth]{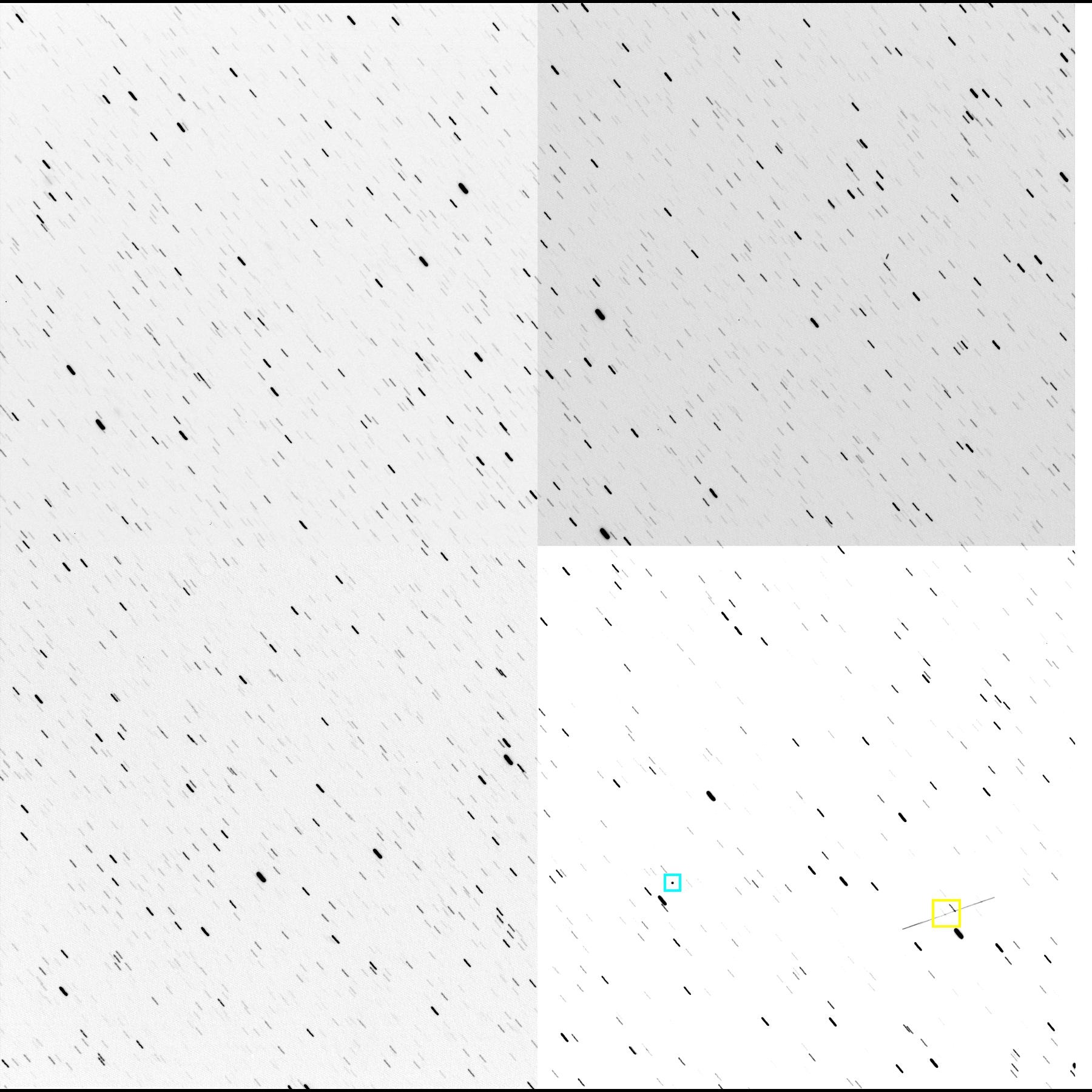} 
    \caption{}  
    \label{fig:fig11b}  
\end{subfigure}  
\begin{subfigure}{0.9\columnwidth}  
    \centering  
    \includegraphics[width=\columnwidth]{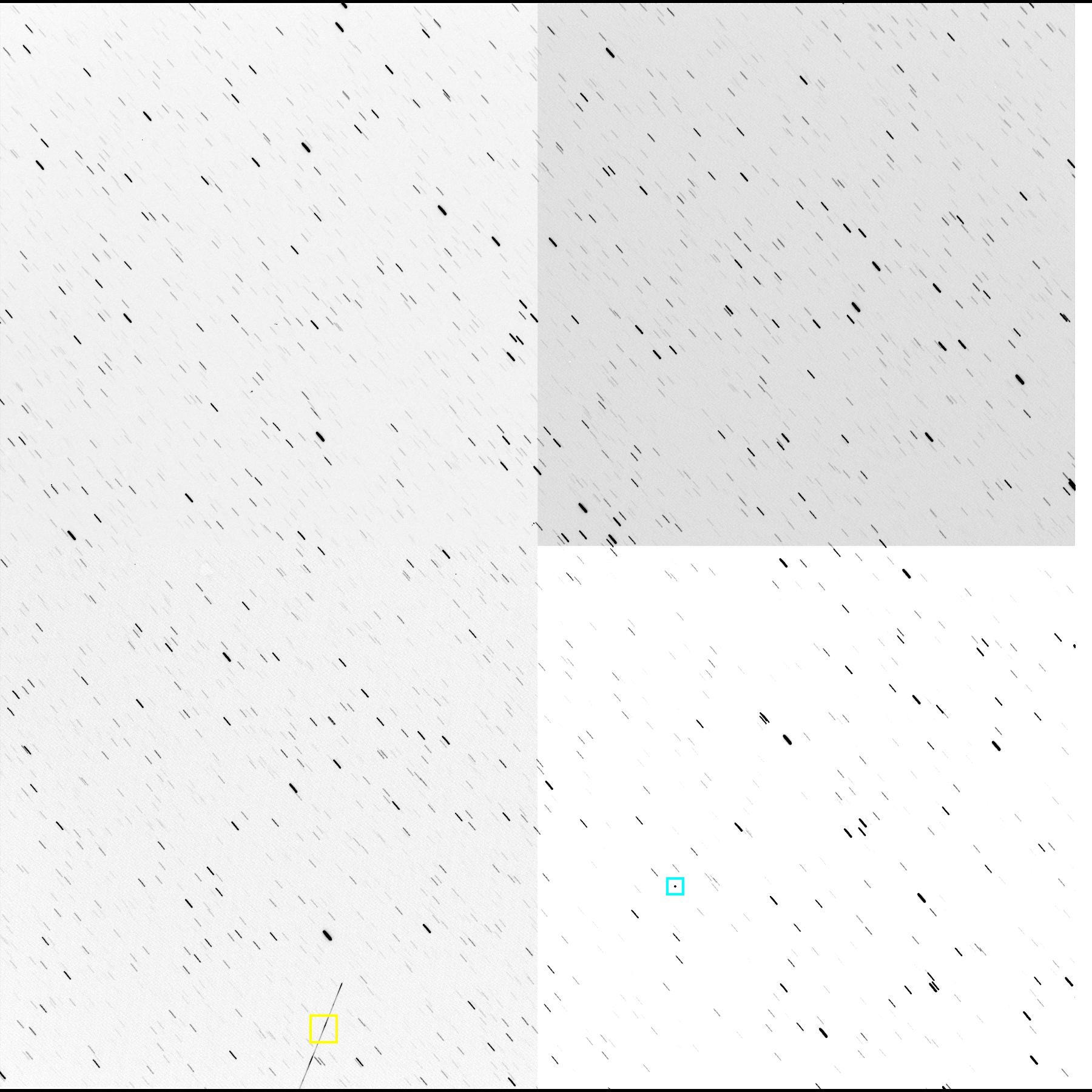} 
    \caption{}  
    \label{fig:fig11b}  
\end{subfigure}  
\begin{subfigure}{0.9\columnwidth}  
    \centering  
    \includegraphics[width=\columnwidth]{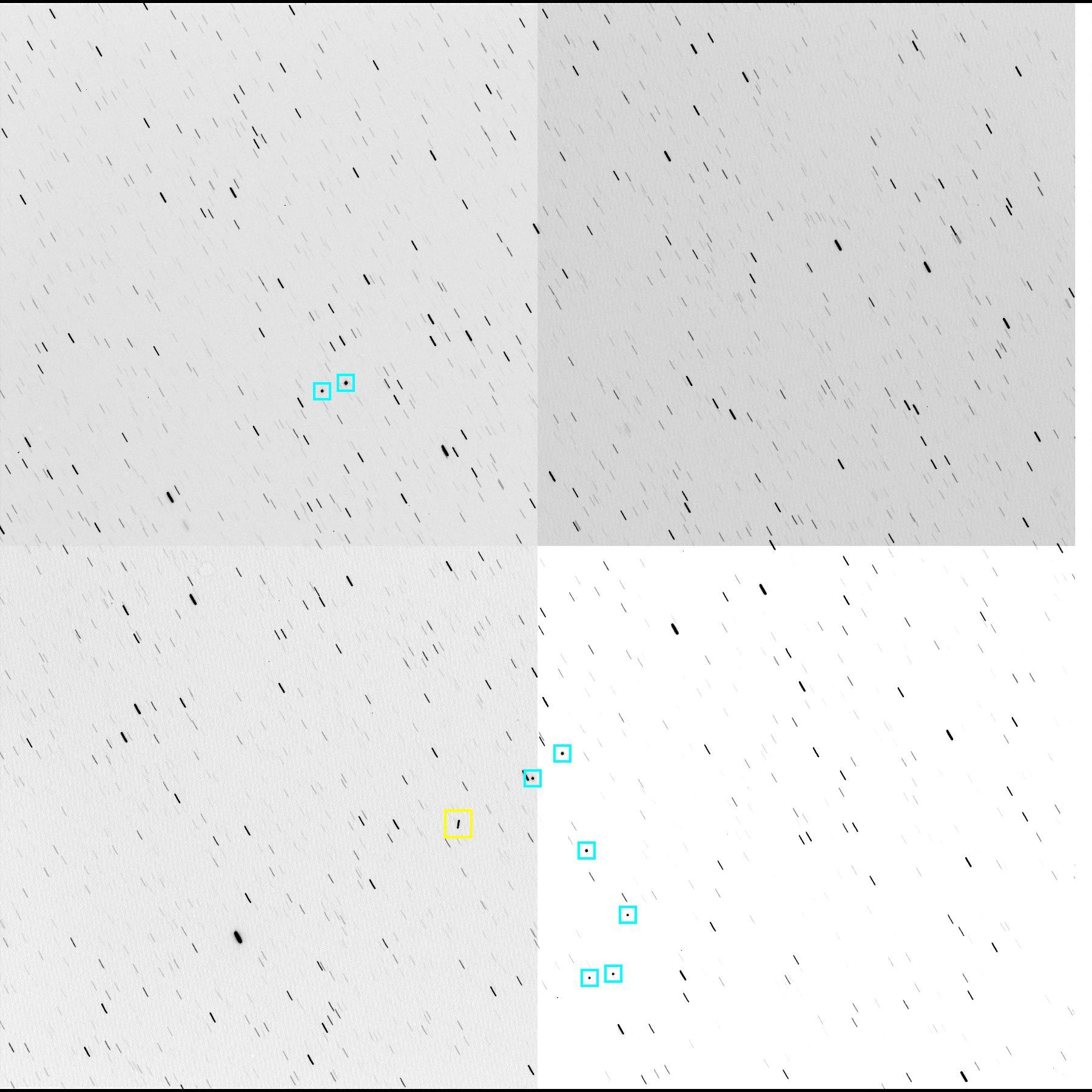} 
    \caption{}  
    \label{fig:fig11b}  
\end{subfigure}  

\caption{The display of the detection result in the original image, where the target in the blue box is the tracked target, and the target in yellow is the target in other motion modes. We performed a black-and-white color inversion on the image. }
\label{fig:fig18}  

\end{figure*}

The detection results of tracked targets and other moving objects obtained by the method in this paper are shown in Table~\ref{tab:8}, and the display of detection results in the original image is shown in Fig.~\ref{fig:fig18}.

\begin{table}  
\renewcommand{\arraystretch}{1}  
\caption{Detection results of all types of moving objects, tracked target (TT). targets with other motion modes (TOM), and others in validation dataset.}  
\label{tab:8}  
\centering  
\begin{tabular}{ccccc}  
\toprule  
& &\multicolumn{3}{c}{Detection} \\  
& Class & TT & TOM & Other \\  
\midrule  
\multirow{3}{*}{Ground Truth}   
 & TT & 239 &	0 &	6\\
 & TOM   & 0 & 30 & 9  \\
 & Other & 0 & 15 & -  \\
\midrule  
\multicolumn{2}{c}{$P_d$} & \multicolumn{3}{c}{94.72 $\% $} \\  
\multicolumn{2}{c}{FAR} & \multicolumn{3}{c}{5.02 $\%$} \\  
\bottomrule  
\end{tabular}  
\end{table}

The application of a neural network model can reduce the manual determination of the threshold range, and obtain better results, and through the major axis length and inclination angle statistics can be more accurate detection of other moving modes of the moving object. With a hardware configuration of an Intel 8-core CPU, 3GB of RAM, and an NVIDIA GeForce GT 710 graphics card, the processing time of Feature Extraction Software for a 4K image feature extraction is about 0.17 seconds/frame, and the detection time of moving object based on feature extraction is 0.49 seconds/frame. The total running time of this method is 0.66 seconds/frame, which meets the real-time processing requirements.

\section{Conclusions}

The paper presents a method for detecting multiple moving objects in wide-field images using non-sidereal tracking.
Firstly, feature information is extracted from optical images, and the stars and tracked targets are automatically labeled. Then, the data is trained by the fully connected neural network model, and the classification results are obtained according to the model, and the tracked targets are detected by KD-tree track association, and other motion modes targets are detected by statistics of inclination angle and major axis length data. The test results of the actual observation images show that the accuracy of the moving object detection can reach 94.72\% and the false alarm rate is 5.02\%. In addition, the method has the following advantages:

(1) The method can not only accurately detect the tracked target in the image and the target with the same speed as the tracked target, but also realize the detection of other targets with moving modes in the image.

(2) The neural network model in this method uses the data extracted from images for training and can obtain good training effects. This demonstrates that the model can be effectively trained without the accumulation of excessive data in the image samples, enabling the algorithm to achieve satisfactory training outcomes even with a limited number of observational images. This method has a small amount of computation, not only can enhance the adaptability of the system, but also can apply the network to real-time data processing.


Our research provides an effective and practical solution for the detection of multiple moving objects in wide-field images with non-sidereal tracking. This method is not only highly flexible and adaptable but also significantly reduces the demand for training data, achieving rapid and precise detection of moving objects. Due to data limitations, this method is only tested in the non-sidereal tracking mode, but theoretically, the process of this method is also applicable to other observation  modes. Looking ahead, we will continue to deepen our research and apply more experimental data to various practical scenarios to validate and optimize our algorithm. At the same time, we will also focus on the impact of overexposed stars in images and the detection of fast-moving objects, and consider using more accurate feature information such as FWHM and more accurate signal-to-noise ratio formulas to improve the accuracy and adaptability of the moving objects detection algorithm.

\section*{Acknowledgements}

Our work was funded by the National Science and Technology Major Project (2022ZD0117401), and the National Natural Science Foundation of China (12273063). We acknowledge the support of the staff of the Nanshan One-meter Wide-field Telescope.












\section*{Data Availability}

The data will be shared on reasonable request to the corresponding author. The software utilized in the present research is proprietary and is not openly accessible to the public. For any collaboration initiatives or inquiries specifically pertaining to the software, kindly address your correspondence to the corresponding author, who will undertake an assessment of the viability of software sharing contingent upon reasonable and mutually beneficial proposals.


\section*{Software}

\begin{itemize}
    \item Astropy \citep{2013A&A...558A..33A,2018AJ....156..123A}
    
    \item Numpy \citep{harris2020array}
    
    \item Scikit-learn \citep{scikit-learn}
    
    \item Pandas \citep{reback2020pandas}

    \item PyTorch \citep{paszke2019pytorch}
    
    \item SciPy \citep{2020SciPy-NMeth}
    
    \item Matplotlib \citep{Hunter}
    
    \item Photutils  \citep{larry_bradley_2023_7946442}
    
    \item Skimage   \citep{van2014scikit}

    \item Math \citep{van1995python}

\end{itemize}



\bibliographystyle{mnras}
\bibliography{example} 








\bsp	
\label{lastpage}
\end{document}